\documentclass[10pt, a4paper]{article}
\usepackage[left=2cm, right=2cm, top=2cm, bottom=2cm, bindingoffset=0cm]{geometry}

\usepackage{cmap} 
\usepackage[utf8x]{inputenc}

\usepackage{amssymb, amsmath, amscd}

\usepackage[bookmarksnumbered=true]{hyperref}
\numberwithin{equation}{section}

\usepackage{tikz}
\usetikzlibrary{cd, arrows, decorations.pathreplacing, decorations.markings}
\usetikzlibrary{calc, math}

\def\Am{\mathcal{A}}
\def\Ron{\mathcal{R}}
\def\prep{\mathcal{F}}
\def\Re{\mathrm{Re}\,}
\def\Im{\mathrm{Im}\,}
\def\Ker{\mathrm{Ker}\,}
\def\Id{\mathrm{Id}\,}
\def\Dom{\Omega}

\def\E{\mathrm{E}}
\def\det{\mathrm{det}\,}

\def\Gr{\Gamma}
\def\Dconf{\mathcal{D}}
\def\Z{\mathcal{Z}}

\def\curve{\mathcal{C}}

\def\Cc#1{C^{#1}(\Gr,\mathbb{R})}
\def\Cct#1{C^{#1}(\Gr_{1},\mathbb{R})}

\def\we#1{\mathrm{#1}}
\def\Rp{\mathbb{R}_{> 0}}
\def\db{\varepsilon}
\def\T{\mathrm{T}}
\def\Tv{\overleftarrow{\T}}
\def\K{\mathrm{K}}
\def\Kc{\K_{1}}
\def\Kct{\tilde{\K}_{1}}
\def\Q{\mathrm{Q}}
\def\G{\mathrm{G}}

\def\M{\mathrm{M}}
\def\ksign#1{(-1)^{\varkappa_{#1}}}

\def\Vol{\mathrm{Vol}}
\def\R5{R_5}

\def\zc{{z^{\vee}}}
\def\wc{{w^{\vee}}}

\usepackage{color}

\def\ii{\mathrm{i}\mkern1mu}

\title{\boldmath Topological string amplitudes and Seiberg-Witten prepotentials \\ from the counting of dimers in transverse flux}

\author{M. Semenyakin\thanks{semenyakinms@gmail.com}}
\date{\small\textit{
Instituut-Lorentz, Universiteit Leiden, P.O. Box 9506, 2300 RA Leiden, The Netherlands
}}

\begin{document}

\tikzset{
	blackCircle/.style={fill=black,thick,radius=0.1,inner sep=0},
	whiteCircle/.style={fill=white,thick,radius=0.1,inner sep=0},
	quiverVertex/.style={fill=black,thick,radius=0.07},
	styleQuiverEdge/.style={
		line width=0.7pt, 
		postaction={decorate},
		decoration={markings,
		mark=at position 0.6 with {\arrow{Stealth[scale=1]}}
		}
	},
	styleArrow/.style 2 args={
		postaction={decorate},
		decoration={markings, mark=at position #1 with {\arrow{Stealth[scale=#2]}}}
		}
}


\maketitle
\begin{abstract}
\noindent
Important illustration to the principle {\it ``partition functions in string theory are $\tau$-functions of integrable equations''} is the fact that the (dual) partition functions of $4d$ $\mathcal{N}=2$ gauge theories solve Painlev\'{e} equations. In this paper we show a road to self-consistent proof of the recently suggested generalization of this correspondence: partition functions of topological string on local Calabi-Yau manifolds solve $q$-difference equations of non-autonomous dynamics of the ``cluster-algebraic''integrable systems.

We explain in details the ``solutions'' side of the proposal. In the simplest non-trivial example we show how $3d$ box-counting of topological string partition function appears from the counting of dimers on bipartite graph with the discrete gauge field of ``flux'' $q$. This is a new form of topological string/spectral theory type correspondence, since the partition function of dimers can be computed as determinant of the linear $q$-difference Kasteleyn operator. Using WKB method in the ``melting'' $q\to 1$ limit we get a closed integral formula for Seiberg-Witten prepotential of the corresponding $5d$ gauge theory. The ``equations'' side of the correspondence remains the intriguing topic for the further studies.

\end{abstract}

\section{Introduction}
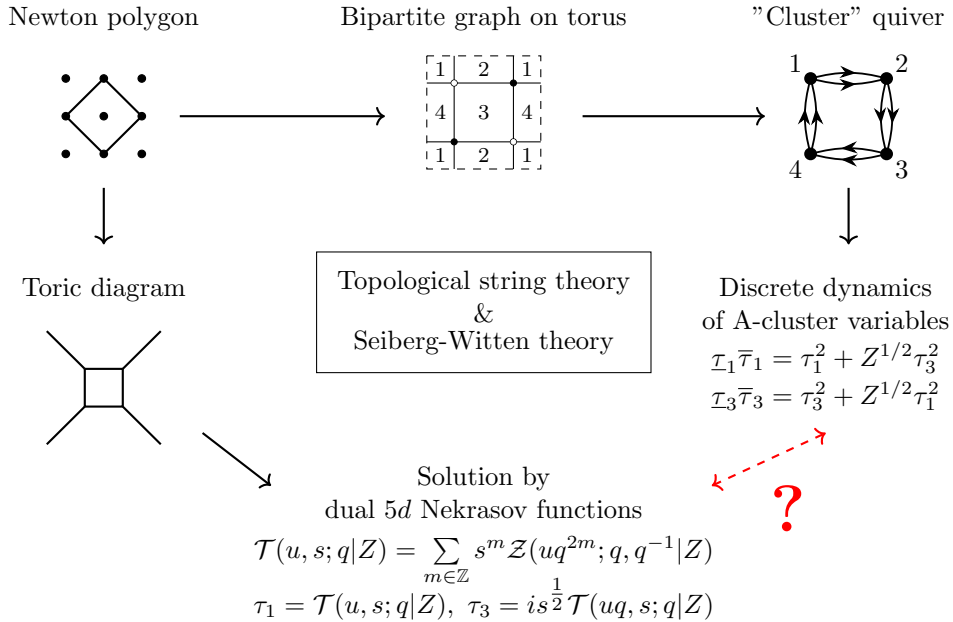
\begin{figure}[!h]
\begin{center}
\begin{tikzpicture}

\tikzmath{\xs=-2.5;\ys=2.2;};
\node at (\xs+0.5,\ys+0.8) {Newton polygon};
\foreach \x in {0,...,2}
	\foreach \y in {0,...,2}
		\draw[fill] (\xs+0.5*\x,\ys-0.5*\y) circle[radius=0.05];

\draw[thick] (\xs,\ys-0.5) -- (\xs+0.5,\ys) -- (\xs+1,\ys-0.5) -- (\xs+0.5,\ys-1) -- (\xs,\ys-0.5);

\draw[thick, ->] (-1,1.7) -- (1.7,1.7);
\draw[thick, ->] (-2,0.75) -- (-2,0);

\tikzmath{\xs=-2.5;\ys=-1.4;};
\node at (\xs+0.5,\ys+0.8) {Toric diagram};

\draw[thick] (\xs+0.25,\ys-0.25) -- (\xs+0.75,\ys-0.25) -- (\xs+0.75,\ys-0.75) -- (\xs+0.25,\ys-0.75) -- (\xs+0.25,\ys-0.25);
\draw[thick] (\xs+0.25,\ys-0.25) -- (\xs-0.25,\ys+0.25);
\draw[thick] (\xs+0.75,\ys-0.25) -- (\xs+1.25,\ys+0.25);
\draw[thick] (\xs+0.75,\ys-0.75) -- (\xs+1.25,\ys-1.25);
\draw[thick] (\xs+0.25,\ys-0.75) -- (\xs-0.25,\ys-1.25);

\draw[thick, ->] (-0.7,-2.5) -- (0.2,-3.2);

\tikzmath{\xs=2.25;\ys=2.5;};
\node at (\xs+0.75,\ys+0.5) {Bipartite graph on torus};
\draw (\xs+0,\ys-1.14) -- (\xs+1.5,\ys-1.14);
\draw (\xs+0,\ys-0.36) -- (\xs+1.5,\ys-0.36);
\draw (\xs+0.36,\ys+0) -- (\xs+0.36,\ys-1.5);
\draw (\xs+1.14,\ys+0) -- (\xs+1.14,\ys-1.5);
\draw[dashed] (\xs,\ys) rectangle (\xs+1.5,\ys-1.5);
\draw[fill] (\xs+0.36,\ys-0.36) circle[radius=0.04];
\draw[fill] (\xs+0.36,\ys-1.14) circle[radius=0.04];
\draw[fill] (\xs+1.14,\ys-0.36) circle[radius=0.04];
\draw[fill] (\xs+1.14,\ys-1.14) circle[radius=0.04];
\draw[fill, white] (\xs+0.36,\ys-0.36) circle[radius=0.03];
\draw[fill, white] (\xs+1.14,\ys-1.14) circle[radius=0.03];

\node at (\xs+0.18,\ys-0.18) {\footnotesize 1};
\node at (\xs+1.32,\ys-0.18) {\footnotesize 1};
\node at (\xs+0.18,\ys-1.32) {\footnotesize 1};
\node at (\xs+1.32,\ys-1.32) {\footnotesize 1};

\node at (\xs+0.75,\ys-0.18) {\footnotesize 2};
\node at (\xs+0.75,\ys-1.32) {\footnotesize 2};

\node at (\xs+0.75,\ys-0.75) {\footnotesize 3};

\node at (\xs+0.18,\ys-0.75) {\footnotesize 4};
\node at (\xs+1.32,\ys-0.75) {\footnotesize 4};

\draw[thick, ->] (4.3,1.7) -- (6.7,1.7);

\tikzmath{\xs=7.3;\ys=1.2;\bend=15;\d=1;};
\node at (\xs+0.5,\ys+1.8) {"Cluster" quiver};
\draw[quiverVertex] (\xs+0,\ys) circle;
\draw[quiverVertex] (\xs+0,\ys+\d) circle;
\draw[quiverVertex] (\xs+\d,\ys+\d) circle;
\draw[quiverVertex] (\xs+\d,\ys) circle;

\node at (\xs-0.2,\ys+\d+0.2) {1};
\node at (\xs+\d+0.2,\ys+\d+0.2) {2};
\node at (\xs+\d+0.2,\ys-0.2) {3};
\node at (\xs-0.2,\ys-0.2) {4};

\path (\xs,\ys) edge[styleQuiverEdge, bend right=\bend] (\xs,\ys+\d);
\path (\xs,\ys) edge[styleQuiverEdge, bend left=\bend] (\xs,\ys+\d);	
	
\path (\xs,\ys+\d) edge[styleQuiverEdge, bend right=\bend] (\xs+\d,\ys+\d);
\path (\xs,\ys+\d) edge[styleQuiverEdge, bend left=\bend] (\xs+\d,\ys+\d);	
	
\path (\xs+\d,\ys+\d) edge[styleQuiverEdge, bend right=\bend] (\xs+\d,\ys);
\path (\xs+\d,\ys+\d) edge[styleQuiverEdge, bend left=\bend] (\xs+\d,\ys);
	
\path (\xs+\d,\ys+0) edge[styleQuiverEdge, bend right=\bend] (\xs,\ys);
\path (\xs+\d,\ys+0) edge[styleQuiverEdge, bend left=\bend] (\xs,\ys);
	
\draw[thick, ->] (7.8,0.75) -- (7.8,0);

\tikzmath{\xs=7.5;\ys=-0.5;};
\node at (\xs,\ys-0.1) {Discrete dynamics};
\node at (\xs,\ys-0.5) {of A-cluster variables};
\node at (\xs,\ys-1) {$\underline{\tau}_1 \overline{\tau}_1 = \tau_1^2+Z^{1/2}\tau_3^2$};
\node at (\xs,\ys-1.5) {$\underline{\tau}_3 \overline{\tau}_3 = \tau_3^2+Z^{1/2}\tau_1^2$};

\draw[thick, dashed, red, <->] (6,-3.2) -- (7.5,-2.5);
\node[red] at (7,-3.5) {{\Huge \textbf{?}}};

\tikzmath{\xs=3;\ys=-3;};
\node at (\xs,\ys-0.1) {Solution by};
\node at (\xs,\ys-0.5) {dual $5d$ Nekrasov functions};
\node at (\xs,\ys-1.2) {$\mathcal{T}(u,s;q|Z) = \sum\limits_{m\in \mathbb{Z}}s^m \mathcal{Z}(uq^{2m};q,q^{-1}|Z)$};
\node at (\xs,\ys-1.7) {$\tau_1 = \mathcal{T}(u,s;q|Z), ~ \tau_3 = i s^{\tfrac{1}{2}}\mathcal{T}(uq,s;q|Z)$};

\draw (0.8,-0.1) rectangle (5.2,-1.7);
\node at (3,-0.5) {Topological string theory};
\node at (3,-0.9) {\&};
\node at (3,-1.3) {Seiberg-Witten theory};

\end{tikzpicture}
\caption{The long way between the $q$-difference equations and their solutions}
\label{fig:mainDiagram}
\end{center}
\end{figure}

\paragraph{Proposal of the paper.} We are going to briefly explain the structure of the puzzle we are aiming to solve, as it is shown on Fig.~\ref{fig:mainDiagram}, and our proposal how it could be solved. Then we will get a little bit more into context and motivation of it.

Starting from Newton polygon, which is convex polygon with integral coordinates of vertices, one can construct two seemingly unrelated objects. Either go down to toric diagram, encoding some toric $3d$ Calabi-Yau manifold, partition function of topological string on which can be computed using topological vertices, and which sometimes coincides with the partition function of instantons of $5d$ $N=1$ gauge theory. We will call ``Fourier'' transorm of it to be dual partition function. Or go to the right of the figure, constructing bipartite graph on torus, which encodes cluster integrable system, whose spectral curve have the same Newton polygon. One can deautonomize this system in a canonical way, loosing involutive and preserved hamiltonians, but getting bilinear $q$-difference equations on $A$-cluster variables, for any element of cluster mapping class group of the quiver. It was checked in some examples \cite{BGM17}, \cite{BGT17}, \cite{BGM18} (all of which were of field-theoretic type), that the string-theoretic partition functions mentioned above are satisfy equations coming from cluster algebras. And despite of the simplicity of formulation, there is no proof for the general Newton polygon yet.

The proposal of this paper is to show how the partition functions of topological string can be obtained in purely cluster algebraic setting, building the missing red arrow on Fig.~\ref{fig:mainDiagram}. We claim that in order to deautonomize the cluster integrable system, one has to uplift the Kasteleyn operator from torus to the plane, covering the torus. The deautonomization parameter $q$ plays a role of the transverse flux of discrete $\mathbb{R}_{>0}$-connection. The partition function of dimers, which provided spectral curve in the autonomous case, becomes a partition function of dimers on the infinite plane. We claim, that being properly regularized and with certain scaling of parameters, this partition function reproduces the counting of topological vertices, which constitute topological string partition function.

This proposal is well agreed with the topological string/spectral theory correspondence like in \cite{BGT17}, since the partition function of dimers on a plane can be computed using the determinant of Kasteleyn operator, which in this case is almost a quantization of spectral curve. Cluster algebraic interpretation of partition functions opens a room for proving bilinear relations among them as for $A$-cluster variables related by mutations of the cluster seed.

\paragraph{Equations and partition functions.} Conjecture that the instanton partition functions solve Painlev\'{e} equations was proposed for the first time in \cite{GIL12}. The motivation for the solution came there from a relation between the theory of isomonodromic deformations and the theory of holonomic fields \cite{SMJ79}. The claim was that the $\tau$-function of Painlev\'{e} VI equation, which encodes isomonodromic deformations of rank two Fuchsian system with four punctures on $\mathbb{C}P^1$, is equal to the chiral correlating function of four generic primary operators in $c=1$ conformal field theory. Using AGT correspondence \cite{AGT09} this function was written there as Fourier transformation of $4d$ $\mathcal{N}=2$ $SU(2)$ gauge theory with $N_f=4$ flavours. The correspondence was immediately generalized by the same authors to the partition functions of theories with $N_f=0,1,2,3$ as solutions to Painlev\'{e} III and V equations in \cite{GIL13}. It was promoted to higher rank \cite{G15}, \cite{GM16}, \cite{GL16}, \cite{GIL18}, with Virasoro algebra being replaced by $W_N$ algebra. In gauge theoretic terms it was shown that the partition function of $SU(N)$ theory with the linear quiver of length $n$ solves isomonodromic equations for the Fuchsian system of rank $N$ with $2$ full and $n-2$ semi-degenerate punctures \cite{GL16}. Important observation was that all of the conformal field theories involved in the correspondence were free-fermionic, so the $\tau$-functions were shown to be free-fermionic then \cite{ILT14}, \cite{GM16}, \cite{GL16}.

Natural deformation of the approach of \cite{GIL12} was to solve $q$-difference Painlev\'{e} equations with the partition functions of $5d$ $\mathcal{N}=1$ supersymmetric theories. The first example of this kind was the solution of $q$-Painlev\'{e} III equation with the partition functions of $5d$ $SU(2)$ gauge theory without matter in \cite{BS16}. More general case of $q$-Painlev\'{e} VI and theories with $N_f = 4$ flavours was considered in \cite{JNS17}. In \cite{BGT17} it was suggested that there should be similar formulas for the solutions of all $q$-Painleve equations, which might be classified using the Newton polygons\footnote{The Newton polygons are just convex polygons on integral plane $\mathbb{Z}^2$, which will play important role in the following.} with one internal point \cite{OY14}, \cite{BGM17}. Since not all of the Newton polygons of this type might be brought into correspondence to some Lagrangian gauge theory with the well-defined partition function of instantons, it was suggested there to use in this case a grand canonical partition function of topological string instead.

The reason for this was that by any Newton polygon one can construct family of $3d$ toric Calabi-Yau manifolds (see e.g. \cite{AKMV03}), and compactification of M-theory on those (or dual $(p,q)$ branes web) defines $5d$ $\mathcal{N}=1$ gauge theory \cite{AHK97}. The striking check of the correspondence between Calabi-Yau manifold and gauge theories was that in the cases when the gauge theory posses Lagrangian description, so that the partition function of instantons in $\Omega$-background can be computed, it can be reproduced by the computation of the partition function of topological string on corresponding manifold \cite{IKP02}, \cite{IKP03}, \cite{EK03}. The computation of the partition function exploited there was based on the ``topological vertex'' technique \cite{AKMV03}: the Calabi-Yau manifolds under consideration are toric, so they can be cutted into $\mathbb{C}^3$ pieces, glued one with another by transition maps. The geometry can be read off from the ``toric diagram'', which is dual as a graph to the triangulated Newton polygon, as on Fig.~\ref{fig:mainDiagram}. Roughly speaking\footnote{Up to subtleties with the ``framing'' and choosing of K\"{a}hler parameters}, to compute the partition function by this picture one associates with each junction of three line segments the topological vertex function
\begin{equation}
V_{\mu ,\nu, \lambda}(q) = \sum_{\pi_{\lambda,\mu,\nu}} q^{|\pi|},
\end{equation} 
which counts $3d$ Young diagrams with $2d$ Young diagrams $\lambda, \mu, \nu$ as asymptotics weighted by the number of boxes in them, and summation over $2d$ Young diagrams weighted by $Q_i$'s to the power of the number of $2d$ boxes to each compact line segment. The parameters $Q_i$ are called K\"{a}hler parameters, and can be treated as exponentiated lengths of the segment on picture, so the parallel segments bounded by the same parallel lines should have equal K\"{a}hler parameters. For the example on Fig.~\ref{fig:mainDiagram} following this rules one gets
\begin{equation}
Z_{\mathrm{boxes}}(q,Q_B, Q_F) = \sum_{\lambda,\mu,\nu,\rho} (Q_B)^{|\lambda|+|\nu|} (Q_F)^{|\mu|+|\rho|} V_{\mu,\nu,\emptyset}(q) V_{\nu,\rho,\emptyset}(q) V_{\rho,\lambda,\emptyset}(q) V_{\lambda,\mu,\emptyset}(q),
\end{equation}
where the empty diagrams are associated with non-compact line segments. Taking summations over $\mu$ and $\rho$, this becomes a partition function of $SU(2)$ theory with no matter multiplets. Being modified by simple ``perturbative'' factor and  ``Fourier'' transformed, as in definition of $\mathcal{T}$ on Fig.~\ref{fig:mainDiagram}, bottom, it becomes a general solution of $q$-Painlev\'{e} III equation shown on Fig.~\ref{fig:mainDiagram}, right, where $\overline{\tau} = \tau(u,s;q|qZ)$, $\underline{\tau} = \tau(u,s;q|q^{-1}Z)$.

Alternative way to compute the same partitions functions utilizes spectral curve \cite{N96}, which is zero locus of Laurent polynomial with the same Newton polygon as those which were used to build the toric Calabi-Yau manifold, and defines its mirror-dual Calabi-Yau manifold. Topological string/spectral theory correspondence states that the partition function of topological string is equal to Fredholm determinant of infinite-dimensional linear operator, quantizing the spectral curve of the system \cite{M15}.

\paragraph{Cluster algebras, dimer models and box-counting.}

The notion of cluster algebras appeared from the solution \cite{FZ98}, \cite{P06} of the total positivity problem of \textit{"How to parametrize all matrices whose minors are strictly positive?"}. The main component of the solution was the certain anzatses for the factorization of matrices, which might be usefully encoded into planar bicoloured graphs with oriented paths on the graphs corresponding to the monomials in parametrization. The weights in the anzatses served as prototypes for $\mathcal{X}$-cluster variables, the minors in the matrices gave birth to $A$-cluster variables, and transformations, identifying equivalent anzatses, became mutations of cluster seeds. The formal definition of cluster algebra was given first in \cite{FZ01}. It appeared soon, that the cluster algebras admit good Poisson structures \cite{GSV02}, \cite{GSV06} which are nicely quantizable \cite{BZ04}, \cite{FG03-2}, and provide a convenient language for parametrization of the spaces of local systems on surfaces \cite{FG03-1}, for theory of stability conditions in algebraic geometry \cite{KS08} and theory of integrable systems \cite{GSV09}, \cite{GK11}, \cite{FM14}.

The important equivalence of counting of paths and counting of dimers on graphs was observed in the context of cluster integrable systems in \cite{GK11},\cite{FM14}. Each configuration of dimers is such set of edges on graph, that each vertex is adjacent to exactly one edge from the set, and the difference of such configurations is always a set of closed paths then. The statistical models of random dimers configurations are well-studied \cite{K67}, \cite{KOS03}, and are free fermionic, and all correlators and partition function of the model might have been written using the minors of Kasteleyn operator, which is basically just the weighted adjacency matrix of the underlying graph. The spectral curve, which is generating function of Hamiltonians of the cluster integrable system was written in \cite{GK11} in the form (\ref{eq:curveToda}), where $\K_1(\lambda,\mu)$ is the Kasteleyn operator of the graph on torus. It was also shown there, how to construct cluster integrable system with the arbitrary Newton polygon of the spectral curve. The coordinates on phase space of cluster integrable system are $\mathcal{X}$-cluster coordinates $x_f$, which can be conveniently interpreted as monodromies of discrete $\mathbb{R}_{>0}$ connection around the faces $f$ of the graph. They are naturally constrained by the condition $\prod_f x_f = 1$ because of the triviality of bundle. In \cite{BGM17} it was shown, that relaxation of the condition to $q\neq 1$ breaks classical integrability of the model, but ``deautonomize'' dynamics generated by the elements of cluster mapping class group. It was also shown there, that $A$-cluster variables provide bilinear form for this dynamics, and for the cases of Newton polygons with one internal point, the corresponding dynamical systems are all $q$-Painlev\'{e} equations except two.

Another appearance of parameter $q$ was in the incarnation of dimer model as a model of statistical physics. The dimer models have nice alternative interpretation as an ensembles of stepped surfaces built from the ``boxes'' having shapes of the faces of graph, which are stacked one on another. The statistical weights of boxes are equal to weights of faces, and for large periodic graph with fixed boundary conditions the flux through the fundamental domain $q$ controls average volume under the surface. The explicit computations of correlating functions for general $q$ were done in \cite{OR01}, \cite{OR05} for hexagonal lattice using the free-fermionic vertex operators, with various boundary conditions. In this case it was just explicitly the problem of the counting of boxes, staying along the wall of the room of complex shape. In the limit $q=e^{-\db} \to 1$ the ``typical'' surface acquires infinite volume $\sim \db^{-3}$. The ``limit shape'' problem of finding its shape were solved first using variational methods in \cite{CKP01} for hexagonal lattice, and then in \cite{KO05} for the general graph and boundary conditions. From the point of view of counting of instantons, the $\db \to 0$ corresponds to Seiberg-Witten limit \cite{NO03}, where the partition function is dominated by single term, with the free-energy density being equal to Seiberg-Witten prepotential of 5d gauge theory \cite{N96}.

Extensive number of attempts were made to connect topological string theory, counting of dimers and cluster algebras in the context of so-called ``crystal melting'' models, see e.g. \cite{ORV03}, \cite{INOV03}, \cite{HV06}, \cite{MN06}, \cite{DOR07}, \cite{Y07}, \cite{OY08}, \cite{MR08}, \cite{CJ08}, \cite{AS10}, \cite{Y10}. The dimer models on bipartite graphs on torus also appeared in string theory in the context of ``brane tiling'' \cite{HK05}, \cite{FHKVW05}, \cite{HV05}, \cite{FHKV05} constructions of $4d$ $\mathcal{N}=1$ theories. Closest to the exposition of this paper consideration were presented in \cite{HKT16}, \cite{HSX17}, \cite{HS20}, where the determinant of tight binding Hamiltonian of particle in magnetic field where attempted to be related to the partition function of topological string at $|q|=1$, and in \cite{KO} where both the ideas of ``transverse magnetic flux'' and of tropicalization were used. Also similar $2d$ lattice operators in the context of the theory of integrable systems were considered e.g. in \cite{K84}, \cite{VKN85}. However, there is yet no consistent proof of the conjecture on how the partition functions of topological string should appear from the counting of dimers on the lattices, built by appropriate Newton polygon.

\paragraph{Structure of the paper.}

In the paper we illustrate all constructions using the single example on Fig.~\ref{fig:mainDiagram}.

In Section \ref{s:dimers} we introduce basic objects and recollect necessary facts on thermodynamic of dimer statistical models. Then we explain how the ``deautonomization'' of $\prod_{f} x_f =q\neq 1$ can be achieved by replacing spectral parameters $\lambda, \mu$ in the Kasteleyn operator of dimers on torus by the $q$-commuting operators of magnetic translations $\tilde{\T}_{x}$, $\tilde{\T}_{y}$. We also discuss degeneracy of their action on the space of functions on $\mathbb{Z}^2$ due to their commutativity with the dual magnetic translations.

In Section \ref{s:WKB} we discuss $q \to 1$ limit. We show how the solution of ``limit shape'' problem can be derived from the WKB approximation for Kasteleyn operator. We show then that the free energy of the model, properly regularized in this limit, gives closed formula for the Seiberg-Witten prepotential of corresponding $5d$ $\mathcal{N}=1$ gauge theory. This also provides regularization for the formula of \cite{OY09} on genus-0 contribution to the partition function of topological string.

In Section \ref{s:boxes} we show how all the necessary box-counting degrees of freedom arise from the counting of dimers, resulting in the main formula of equality of partition function of dimers (in the proper limit) to the dual partition function of topological string
\begin{equation}
\label{eq:ZinstD}
\Z(Q_0 = q, Q_B, Q_F, Q_2) = \sum_{n \in \mathbb{Z}} (Q_2)^{n-1} (Q_B Q_F)^{n(n-1)} q^{\frac{2}{3}n(n-1)(2n-1)} Z_{\mathrm{boxes}}(q, q^{2n} Q_B , q^{2n} Q_F).
\end{equation}
Then, we discuss some issues of inconsistency of the requirements of ``infinite distance'' between the walls of the room, and of ``freezing out'' of non-boxcounting degrees of freedom.

In Section \ref{s:discussion} we outline results of the paper, and propose some directions for the future developments.

\section{Kasteleyn operator of dimers in transverse flux}
\label{s:dimers}

In this section we will show, how making edge weights linearly dependent on the position of fundamental cell, one can relax condition $\prod_{f\in F_1} x_f = q = 1$, deautonomizing cluster integrable system.

\subsection{Zero flux}

\begin{figure}[!h]
\begin{center}
\scalebox{0.98}{\begin{tikzpicture}
\tikzmath{
	\fontsize = 1;
	\fontvert = 0.8;
}

\begin{scope}
\clip(-0.3,-0.3) rectangle (7,7);

\draw (-1,-2)--(-1,6);
\draw (1,-2)--(1,6);
\draw (3,-2)--(3,6);
\draw (5,-2)--(5,6);

\draw (-2,-1)--(6.3,-1);
\draw (-2,1)--(6.3,1);
\draw (-2,3)--(6.3,3);
\draw (-2,5)--(6.3,5);

\draw[blackCircle] (-1,-1) circle;
\draw[blackCircle] (3,-1) circle;
\draw[blackCircle] (1,1) circle;
\draw[blackCircle] (5,1) circle;
\draw[blackCircle] (3,3) circle;
\draw[blackCircle] (-1,3) circle;
\draw[blackCircle] (1,5) circle;
\draw[blackCircle] (5,5) circle;

\draw[whiteCircle] (1,-1) circle;
\draw[whiteCircle] (5,-1) circle;
\draw[whiteCircle] (-1,1) circle;
\draw[whiteCircle] (3,1) circle;
\draw[whiteCircle] (1,3) circle;
\draw[whiteCircle] (5,3) circle;
\draw[whiteCircle] (-1,5) circle;
\draw[whiteCircle] (3,5) circle;

\node[font=\bfseries, scale=\fontsize] at (0, 6) {$4$};
\node[font=\bfseries, scale=\fontsize] at (2, 6) {$3$};
\node[font=\bfseries, scale=\fontsize] at (4, 6) {$4$};
\node[font=\bfseries, scale=\fontsize] at (6, 6) {$3$};

\node[font=\bfseries, scale=\fontsize] at (0, 4) {$1$};
\node[font=\bfseries, scale=\fontsize] at (2, 4) {$2$};
\node[font=\bfseries, scale=\fontsize] at (4, 4) {$1$};
\node[font=\bfseries, scale=\fontsize] at (6, 4) {$2$};

\node[font=\bfseries, scale=\fontsize] at (0, 2) {$4$};
\node[font=\bfseries, scale=\fontsize] at (2, 2) {$3$};
\node[font=\bfseries, scale=\fontsize] at (4, 2) {$4$};
\node[font=\bfseries, scale=\fontsize] at (6, 2) {$3$};

\node[font=\bfseries, scale=\fontsize] at (0, 0) {$1$};
\node[font=\bfseries, scale=\fontsize] at (2, 0) {$2$};
\node[font=\bfseries, scale=\fontsize] at (4, 0) {$1$};
\node[font=\bfseries, scale=\fontsize] at (6, 0) {$2$};

\node[font=\bfseries, scale=\fontvert] at (1.5,0.75) {$1,(i,j)$};
\node[font=\bfseries, scale=\fontvert] at (1.5,2.75) {$1,(i,j)$};
\node[font=\bfseries, scale=\fontvert] at (1.7,4.75) {$1,(i,j{+}1)$};
\node[font=\bfseries, scale=\fontvert] at (5.7,0.75) {$1,(i{+}1,j)$};
\node[font=\bfseries, scale=\fontvert] at (5.7,2.75) {$1,(i{+}1,j)$};
\node[font=\bfseries, scale=\fontvert] at (5.9,4.75) {$1,(i{+}1,j{+}1)$};

\node[font=\bfseries, scale=\fontvert] at (3.5,0.75) {$2,(i,j)$};
\node[font=\bfseries, scale=\fontvert] at (3.5,2.75) {$2,(i,j)$};
\node[font=\bfseries, scale=\fontvert] at (3.7,4.75) {$2,(i,j{+}1)$};

\draw[dotted, thick, styleArrow={0.3}{1}] (0.5,-2) -- (0.5,6);
\draw[dotted, thick, styleArrow={0.3}{1}] (4.5,-2) -- (4.5,6);
\draw[dotted, thick, styleArrow={0.3}{1}] (-2,0.5) -- (6,0.5);
\draw[dotted, thick, styleArrow={0.3}{1}] (-2,4.5) -- (6,4.5);

\node[scale=\fontsize] at (1.3,1.95) {$w_1$};
\node[scale=\fontsize] at (5.3,1.95) {$w_1$};
\node[scale=\fontsize] at (1.3,5.85) {$w_1$};
\node[scale=\fontsize] at (5.3,5.85) {$w_1$};

\node[scale=\fontsize] at (0.2,1.2) {$w_2$};
\node[scale=\fontsize] at (4.2,1.2) {$w_2$};
\node[scale=\fontsize] at (0.2,5.2) {$w_2$};
\node[scale=\fontsize] at (4.2,5.2) {$w_2$};

\node[scale=\fontsize] at (1.3,0) {$w_3$};
\node[scale=\fontsize] at (5.3,0) {$w_3$};
\node[scale=\fontsize] at (1.3,4) {$w_3$};
\node[scale=\fontsize] at (5.3,4) {$w_3$};

\node[scale=\fontsize] at (2,1.2) {$w_4$};
\node[scale=\fontsize] at (6,1.2) {$w_4$};
\node[scale=\fontsize] at (2,5.2) {$w_4$};
\node[scale=\fontsize] at (6,5.2) {$w_4$};

\node[scale=\fontsize] at (3.3,4) {$w_5$};
\node[scale=\fontsize] at (3.3,0) {$w_5$};

\node[scale=\fontsize] at (2,3.2) {$w_6$};
\node[scale=\fontsize] at (6,3.2) {$w_6$};

\node[scale=\fontsize] at (3.3,1.95) {$w_7$};
\node[scale=\fontsize] at (3.3,5.85) {$w_7$};

\node[scale=\fontsize] at (4.2,3.2) {$w_8$};
\node[scale=\fontsize] at (0.2,3.2) {$w_8$};

\end{scope}

\node[font=\bfseries, scale=\fontsize] at (0.65,-0.5) {$\gamma_{v,i}$};
\node[font=\bfseries, scale=\fontsize] at (4.7,-0.5) {$\gamma_{v,i{+}1}$};
\node[font=\bfseries, scale=\fontsize] at (-0.2,0.7) {$\gamma_{h,j}$};
\node[font=\bfseries, scale=\fontsize] at (0,4.7) {$\gamma_{h,j{+}1}$};

\begin{scope}[shift={(7.5,0)}]

\clip(-0.3,-0.3) rectangle (6.7,6);

\draw (-1,-2)--(-1,6);
\draw (1,-2)--(1,6);
\draw (3,-2)--(3,6);
\draw (5,-2)--(5,6);

\draw (-2,-1)--(6.3,-1);
\draw (-2,1)--(6.3,1);
\draw (-2,3)--(6.3,3);
\draw (-2,5)--(6.3,5);

\draw[blackCircle] (-1,-1) circle;
\draw[blackCircle] (3,-1) circle;
\draw[blackCircle] (1,1) circle;
\draw[blackCircle] (5,1) circle;
\draw[blackCircle] (3,3) circle;
\draw[blackCircle] (-1,3) circle;
\draw[blackCircle] (1,5) circle;
\draw[blackCircle] (5,5) circle;

\draw[whiteCircle] (1,-1) circle;
\draw[whiteCircle] (5,-1) circle;
\draw[whiteCircle] (-1,1) circle;
\draw[whiteCircle] (3,1) circle;
\draw[whiteCircle] (1,3) circle;
\draw[whiteCircle] (5,3) circle;
\draw[whiteCircle] (-1,5) circle;
\draw[whiteCircle] (3,5) circle;

\node[scale=\fontsize] at (1.3,1.95) {$w_1$};
\node[scale=\fontsize] at (5.3,1.95) {$w_1$};
\node[scale=\fontsize] at (1.3,5.85) {$w_1$};
\node[scale=\fontsize] at (5.3,5.85) {$w_1$};
\node[scale=\fontsize] at (0.3,1.3) {$w_2 q^{\frac{j}{2}}$};
\node[scale=\fontsize] at (4.1,1.3) {$w_2 q^{\frac{j}{2}}$};
\node[scale=\fontsize] at (0.35,5.3) {$w_2 q^{\frac{j{+}1}{2}}$};
\node[scale=\fontsize] at (4.2,5.3) {$w_2 q^{\frac{j{+}1}{2}}$};
\node[scale=\fontsize] at (1.6,0) {$w_3 q^{{-}\frac{i}{2}}$};
\node[scale=\fontsize] at (5.75,0) {$w_3 q^{{-}\frac{i{+}1}{2}}$};
\node[scale=\fontsize] at (1.6,4) {$w_3 q^{{-}\frac{i}{2}}$};
\node[scale=\fontsize] at (5.75,4) {$w_3 q^{{-}\frac{i{+}1}{2}}$ };
\node[scale=\fontsize] at (2,1.2) {$w_4$};
\node[scale=\fontsize] at (6,1.2) {$w_4$};
\node[scale=\fontsize] at (2,5.2) {$w_4$};
\node[scale=\fontsize] at (6,5.2) {$w_4$};

\node[scale=\fontsize] at (3.5,4) {$w_5 q^{\frac{i}{2}}$};
\node[scale=\fontsize] at (3.5,0) {$w_5 q^{\frac{i}{2}}$};
\node[scale=\fontsize] at (2,3.2) {$w_6$};
\node[scale=\fontsize] at (6,3.2) {$w_6$};
\node[scale=\fontsize] at (3.3,1.95) {$w_7$};
\node[scale=\fontsize] at (3.3,5.85) {$w_7$};
\node[scale=\fontsize] at (4.1,3.3) {$w_8 q^{{-}\frac{j}{2}}$};
\node[scale=\fontsize] at (0.3,3.3) {$w_8 q^{{-}\frac{j}{2}}$};

\end{scope}

\begin{scope}[shift={(15.5,0)}]

\node[scale=\fontsize] at (0.09,5) {$\tilde{x}_1 = q \dfrac{w_2 w_8}{w_3 w_5}$};
\node[scale=\fontsize] at (0,4) {$\tilde{x}_2 = \dfrac{w_3 w_5}{w_4 w_6}$};
\node[scale=\fontsize] at (0,3) {$\tilde{x}_3 = \dfrac{w_4 w_6}{w_1 w_7}$};
\node[scale=\fontsize] at (0,2) {$\tilde{x}_4 = \dfrac{w_1 w_7}{w_2 w_8}$};
\node[scale=\fontsize] at (0,1) {$\tilde{x}_1 \tilde{x}_2 \tilde{x}_3 \tilde{x}_4 = q$};
\end{scope}

\end{tikzpicture}}

\caption{Example of bipartite graph, known to describe Toda integrable chain on two sites. Left: labelling of vertices, faces, and edge weights. Since we consider only periodic weightings of faces, we do not put labels of their fundamental domains on the plot. Right: edges weighting of finite flux $q=e^{-\db}$, according to (\ref{eq:qweight}) and face weights expressed in terms of edge weights.}
\label{fig:todanetwork}
\end{center}
\end{figure}
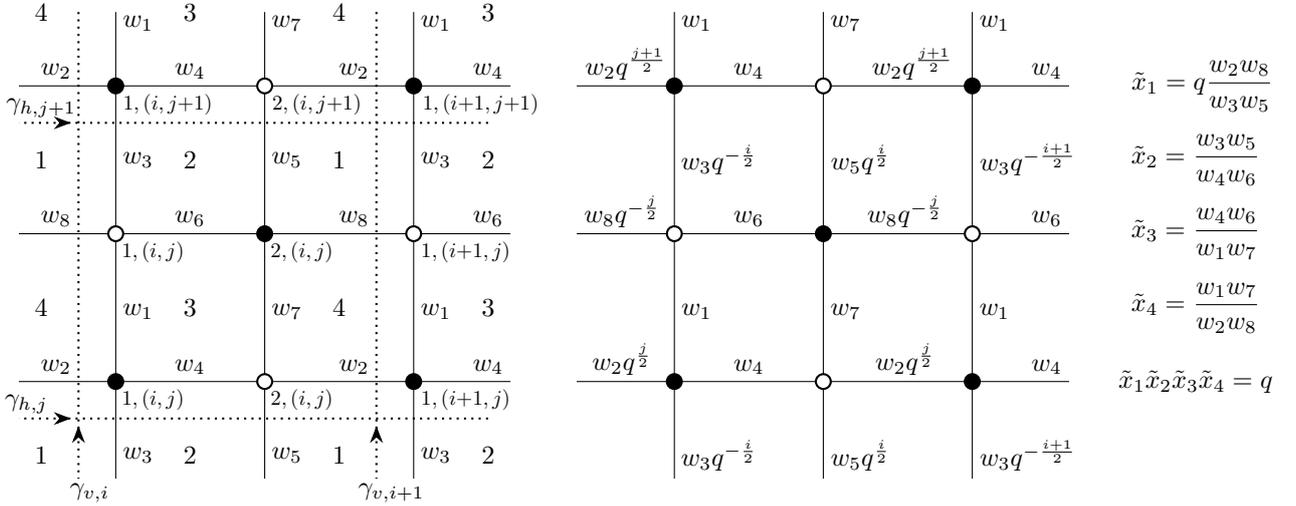

\paragraph{Definition of the model.} The dimer models are usually defined on bipartite graphs, such graphs $\Gr$ that the vertices $V$ can be decomposed into black and white subsets $V = B \sqcup W$, and edges connect only vertices of the opposite colours, see example of Fig.~\ref{fig:todanetwork}. Throughout the paper we assume the graphs to be minimal in the sense of \cite{GK11}. The edges $e\in E$ are weighted by the positive real statistical weights $w_e \in \Rp$ for edges oriented from black to white vertex (which is assumed to be canonical in the following), and by weights $w_{-e} = w_e^{-1}$ for the edges taken with opposite orientations. We also extend multiplicatively $w$ to any sets $S$ of edges by $w_{S} = \prod_{e\in S} w_e$. It is often instructive to consider edge weights as discrete connections in $\Rp$-bundle over $\Gr$.

The possible microscopic states of the model are dimers configurations $D\in \Dconf(\Gr)$ (also called perfect matchings) on $\Gr$, which are such collections of edges of $\Gr$, that each vertex have exactly one adjacent edge from this collection and all edges are taken with the canonical black-to-white orientation. The partition function can be defined, as usual, as a sum of statistical weights over all configurations 
\begin{equation}
\Z(\Gr,w) = \sum_{D\in \Dconf(\Gr)} w_{D}.
\end{equation}
It changes by simple common factor $\Z(\Gr,w) \mapsto \left(\prod_{v\in B} g_v^{-1} \right)\left(\prod_{v\in W} g_v \right)\Z(\Gr,w)$ under $\Rp$ gauge transformations of edge weights
\begin{equation}
w_e \mapsto g_{t(e)} w_e g_{s(e)}^{-1}
\end{equation}
where $g$ is $\Rp$-valued function on vertices, and $s(e)$, $t(e)$ are starting and terminal vertices of edge $e$. So it is meaningful to consider the partition function normalized by the weight of some fixed dimers configuration $D_0$
\begin{equation}
\label{eq:partfnorm}
\Z(\Gr,w;D_0) = \dfrac{\Z(\Gr,w)}{w_{D_0}} = \sum_{D\in \Dconf(\Gr)} w_{D-D_0}.
\end{equation} 
which depends, for planar graphs, only on gauge invariant face weights $x_f = \prod_{e \in \partial f} w_e$, since for any dimers configurations $D,D_0$ holds $\partial(D-D_0) = 0$ and any cycle in a disk is contractible.

\paragraph{Kasteleyn operator.} The dimer models are ``free fermionic'': it simply follows from the definition of determinant, that their partition functions can be effectively computed \cite{K67} as determinants 
\begin{equation}
\label{eq:ZbyK}
\Z(\Gr,w) = \pm \det \K_{\Gr}
\end{equation}
where Kasteleyn matrix $\K_{\Gr}: \mathbb{C}^{|B|} \to \mathbb{C}^{|W|}$ is twisted by additional signs weighted adjacency matrix of $\Gr$
\begin{equation}
(\K_{\Gr})_{\alpha,\beta} = \sum_{\partial e = \alpha - \beta	} \ksign{e} w_{e},
~~~
\alpha \in W, ~ \beta \in B,
\end{equation}
and signs $\ksign{e}$, called Kasteleyn orientation, for every face $f$ are required to satisfy condition
\begin{equation}
\prod_{e\in \partial f}\ksign{e} = (-1)^{|\partial f|/2+1}.
\end{equation}
For planar graph all Kasteleyn orientations are equivalent up to $\mathbb{Z}/2\mathbb{Z}$ gauge transformations
\begin{equation}
\ksign{e} \mapsto (-1)^{\sigma_{s(e)} + \sigma_{t(e)}}\, \ksign{e}
\end{equation}
where $(-1)^{\sigma}$ is $\pm 1$-valued function on vertices. The overall sign $\pm$ in (\ref{eq:ZbyK}) is gauge-dependent.

\paragraph{Fugacities of the translation invariant model on infinite lattice.} The bipartite graph is called periodic and planar if it can be embedded into plane $\mathbb{R}^2$ without intersections of edges and in a way invariant under the action of a $\mathbb{Z}^2$ lattice generated by the pair of discrete translations $\T_x,\T_y$. The fundamental domains of this action are cells of rectangular grid, formed by infinite simple horizontal and vertical curves $\gamma_{h,j} = (\T_y)^j \gamma_{h,0}$ and $\gamma_{v,i} = (\T_x)^i \gamma_{v,0}$ transversal to edges, cell $(i,j)$ is bounded by the curves $\gamma_{v,i},\gamma_{v,i+1}$ and $\gamma_{h,i},\gamma_{h,i+1}$, see Fig.~\ref{fig:todanetwork}, left. We decompose set of vertices as $V = V_1 \times \mathbb{Z}^2$, where the first multiplier is finite and counts vertices inside of the cell, and the second denotes position of fundamental cell which a vertex belongs to. We assume that $V_1$ contains equal number of black and white vertices $B_1$ and $W_1$. Sets of edges and faces could be decomposed in a similar way $E = E_1 \times \mathbb{Z}^2$, $F = F_1 \times \mathbb{Z}^2$, where we attribute an edge to the fundamental cell according to the position of the black vertex adjacent to it, and a face intersecting few cells to one of the fundamental cells which it intersects.

If the weighting on periodic graph is also periodic $w_e = w_{\T_x e} = w_{\T_y e}$, then by factorization of plane by $\mathbb{Z}^2$ action we obtain associated model on graph $\Gr_{1}$ embedded in torus $\mathbb{T}^2$, with the sets of vertices, edges and faces $V_1$, $E_1$ and $F_1$, and $\gamma_{h,i}$, $\gamma_{v,j}$ projected to cycles $\gamma_h$, $\gamma_v$ generating $H_1(\mathbb{T}^2,\mathbb{Z})$. Since any closed cycle $p$ on $\Gr_1$ can be decomposed as
\begin{equation}
\label{eq:Gr1homologie}
p = n p_h + m p_v + \sum_{f\in F_1} n_f \partial f
\end{equation}
where $p_h$, $p_v$ are some cycles on $\Gr_1$ homotopic to $\gamma_h$ and $\gamma_v$, the set of gauge-invariant functions on the space of edge weights is generated by face weights $x_f$ and pair of ``monodromies'' $x_h = \prod_{e\in p_h} w_e$, $x_v = \prod_{e\in p_v} w_e$. The face weights of dimer model on torus are not independent, they always satisfy a ``vanishing of total transverse flux'' constraint
\begin{equation}
q = \prod_{f\in F_1} x_f = 1
\end{equation}
since $\sum_{f\in F_1}\partial f = 0$. We will construct the weighting for the model with non-vanishing flux $q$ in the next paragraph. Also, there is no canonical choosing for cycles $p_h, p_v$, however there is a ``twist'' of edge weights by $e^{B_x},e^{B_y} \in \Rp$
\begin{equation}
\label{eq:wtwist}
w_e \mapsto e^{\langle e, \gamma_{v}\rangle B_x + \langle \gamma_{h}, e\rangle B_y } \, w_e
\end{equation}
where $\langle ~, ~ \rangle$ is a skew-symmetric intersection form with the orientation fixed by $\langle \gamma_{h}, \gamma_{v} \rangle = 1$, which do not change face weights, but shifts $x_h \mapsto e^{B_x} x_h$, $x_v \mapsto e^{B_y} x_v$. We will be using $x_f,~ f \in F_1$ and $e^{B_x},e^{B_y}$ as a full set of fugacities, determining model with the vanishing flux.

\subsection{Non-vanishing flux}
Below we will use the additive notations for gauge transformations, edge and face weights
\begin{equation}
g_v = e^{\we{g}_v},
~~~
w_e = e^{\we{w}_e},
~~~
x_f = e^{\we{x}_f},
\end{equation}
where $\we{g}$, $\we{w}$ and $\we{x}$ are cochains from the discrete de Rham complex
\begin{equation}
\begin{tikzcd}
0 \arrow{r} &
\Cc{0} \arrow[r, "d_0"] &
\Cc{1} \arrow[r, "d_1"] &
\Cc{2} \arrow{r} &
0 
\end{tikzcd}
\end{equation}
with the differentials
\begin{equation}
(d_0\we{g})(e) = \mathrm{g}_{t(e)} - \mathrm{g}_{s(e)},
~~~
(d_1\we{w})(f) = \sum_{e \in \partial f} \mathrm{w}_e.
\end{equation}
Using these differentials the gauge transformations and fluxes can be written as 
\begin{equation}
\we{w} \mapsto \we{w} + d_0\we{g}
~~~\text{and}~~~
\we{x}_f = (d_1 \we{w})(f).
\end{equation}
We will also refer to elements of $\Cc{2}$ which are not necessary exact as to face weightings. The classification of discrete $\Rp$-connections on $V$ with arbitrary translation invariant fluxes is provided by the following:

\paragraph{Lemma.}{Choose any face weighting $\tilde{\we{x}}$ on periodic graph, which is translation invariant $\T_{x,y} \tilde{\we{x}} = \tilde{\we{x}}$. Denote total flux through the fundamental cell by $-\db = \sum_{f\in F_1} \tilde{\we{x}}_f$ and fix decomposition
\begin{equation}
\label{eq:qdecomp}
\tilde{\we{x}}_f = \we{x}_f - \left(\chi_f + \sum_{(i,j) \in \mathbb{Z}^2}\delta_{f,f^{\times}_{(i,j)}}\right) \db
\end{equation}
where $\we{x}, \chi$ are translation invariant face weightings of zero flux through the fundamental cell
\begin{equation}
\label{eq:fluxvanish}
\sum_{f\in F_1} \we{x}_f = \sum_{f\in F_1} \chi_f = 0,
\end{equation}
face $f^{\times}_{(i,j)}$ is the face, which the crossing $\gamma_{h,j} \cap \gamma_{v,i}$ belongs to, and $\delta_{f,f'} = 1$ if $f=f'$, and $\delta_{f,f'}=0$ otherwise. Then there is a unique up to gauge transformation discrete connection $\tilde{\we{w}}$ such that $d_1 \tilde{\we{w}} = \tilde{\we{x}}$, and its gauge equivalence class is presented by edge weighting
\begin{equation}
\label{eq:qweight}
\tilde{\we{w}}_{e} = \we{w}_e - \left(\omega_e + \frac{1}{2}\sum_{(i,j)\in \mathbb{Z}^2} i \langle \gamma^{[i,i+1]}_{h,j}, e \rangle + j \langle \gamma^{[j,j+1]}_{v,i}, e \rangle \right)\db
\end{equation}
where $\we{w}$ and $\omega$ are translation invariant edge weightings with fluxes $d_1\we{w} = \we{x}$, $d_1 \omega = \chi$, $\gamma^{[i,i+1]}_{h,j}$ and $\gamma^{[j,j+1]}_{v,i}$ are intervals of $\gamma_{h,j}$ and $\gamma_{v,i}$ bounded by $\gamma_{v,i},\gamma_{v,i+1}$ and $\gamma_{h,j},\gamma_{h,j+1}$ respectively.}

\paragraph{Remark.}{The illustrating example to this Lemma can be found in Fig.~\ref{fig:todanetwork}, right. Note, that we separated part of face weighting of zero total flux into $\we{x}$ and $\chi$, in order to fix fluxes in $\db \to 0$ limit by $\we{x}$ and to control 'direction' along which the total flux vanishes by $\chi$. We also put sign ``$-$'' at $\db$ to have $q<1$ for exponentiated flux $q= e^{-\db}$ at positive values of $\db$.}

\paragraph{Proof.}{To prove existence of $\we{w}$ and $\omega$, push translation invariant fluxes $\we{x}$ and $\chi$ down to $\Gr_{1}$. The conditions that $\we{x},\chi\in \Im d_1$ are equivalent there to $\we{x},\chi \perp \Ker \delta_2$ where codifferential $\delta_2: \Cct{2} \to \Cct{1}$ is defined by
\begin{equation}
(d_1 \we{w}, \we{x})_2 = (\we{w}, \delta_2 \we{x})_1
~~~
\text{with pairings}
~~~
(\we{w}', \we{w}'')_1 = \sum_{e\in E_1} \we{w}'_e \we{w}''_e,
~~~
(\we{x}', \we{x}'')_2 = \sum_{f\in F_1} \we{x}'_f \we{x}''_f,
\end{equation} 
or explicitly by
\begin{equation}
(\delta_2 \we{x})(e) = \we{x}_{t(e*)} - \we{x}_{s(e*)},
\end{equation}
where $e^*$ is the edge of dual graph, obtained from $e$ by counter-clockwise rotation by $90^{\circ}$. Space $\Ker \delta_2$ is one-dimensional and generated by the constant function $\Omega: ~ \Omega_f = 1 ~ \forall ~ f \in F_1$, so orthogonalities $(\we{x},\Omega)_2 = 0 $ and $(\chi, \Omega)_2 = 0$ are guaranteed by (\ref{eq:fluxvanish}).

The $i$ and $j$ depending terms in (\ref{eq:qweight}) contribute to (\ref{eq:qdecomp}) with $- \varepsilon \cdot \delta_{f,f^{\times}_{(i,j)}}$, and generate total flux $-\db$. This can be computed in any example, and then checked that upon adding vertices to $\partial f^{\times}_{(i,j)}$ and moving them in a way, which keeps $\gamma_{h,j} \cap \gamma_{v,i}$ inside of $f^{\times}_{(i,j)}$ and do not put other intersection points inside of it, flux remains the same. Intersections of boundaries of other faces with $\gamma_{h,\bullet}$ and $\gamma_{v,\bullet}$ come in pairs, whose contributions from these terms cancel each other.

To show uniqueness of the gauge orbit, take difference of any pair of discrete connections $\we{w}_0 = \tilde{\we{w}}' - \tilde{\we{w}}''$ both having flux $\tilde{\we{x}}$. It is closed $d_1 \we{w}_0 = 0$ and exact
\begin{equation}
\we{w}_0 = d_0 \we{g},
~~~ 
\we{g}_v = \sum_{e\in p_{v_0, v}} (\we{w}_0)_e,
\end{equation}
where $p_{v_0, v}$ is any path connecting some fixed vertex $v_0$ with $v$, and the sum is path independent as $\sum_{e\in p} (\we{w}_0)_e = 0$ for any closed path $p$, so $\we{g}$ is well defined. Thus, $\we{g}$ provides desired gauge transformation $\tilde{\we{w}}' = \tilde{\we{w}}'' + d_0 \we{g}$. $\hfill \blacksquare$}\\

The Kasteleyn operator $\tilde{\K}: \mathbb{C}^{|B_1|}\otimes \mathbb{C}^{|\mathbb{Z}^2|} \to \mathbb{C}^{|W_1|}\otimes \mathbb{C}^{|\mathbb{Z}^2|}$  constructed from weighting (\ref{eq:qweight}) can be compactly written in terms of $\Gr_{1}$ as
\begin{equation}
\label{eq:kastblock}
\tilde{\K} = \Kct (\tilde{\T}_x, \tilde{\T}_y) = \sum_{e\in E_1} \ksign{e} q^{\omega_e} w_e \cdot E_{t(e),s(e)} \otimes \Tv(e) 
\end{equation}
where $q = e^{-\db}$ is exponentiated flux per fundamental cell, and the translation operator $\Tv(e)$ is ordered along the edge $e$ product over its intersections with $\gamma_h$, $\gamma_v$, which are images of $\gamma_{h,\bullet}$, $\gamma_{v,\bullet}$ under projection from $\mathbb{R}^2$ to $\mathbb{T}^2$
\begin{equation}
\Tv(e) = \underset{p\in e \cap \gamma_{h,v}}{\overleftarrow{\prod}} \left(\tilde{\T}_{x}\right)^{\langle e, \gamma_v \rangle_p} \left(\tilde{\T}_{y}\right)^{\langle \gamma_h, e \rangle_p}
\end{equation}
of the basic $q$-commuting ``magnetic translations'' $\tilde{\T}_{x,y}: \mathbb{C}^{|\mathbb{Z}^2|} \to \mathbb{C}^{|\mathbb{Z}^2|}$
\begin{equation}
\tilde{\T}_x = \sum_{(i,j)\in\mathbb{Z}^2} q^{-\frac{1}{2}j} \, \E_{i+1,i} \otimes \E_{j,j} , ~~~
\tilde{\T}_y = \sum_{(i,j)\in\mathbb{Z}^2} q^{\frac{1}{2}i} \, \E_{i,i} \otimes \E_{j+1,j}, ~~~
\tilde{\T}_y \tilde{\T}_x = q \tilde{\T}_x \tilde{\T}_y.
\end{equation}
The notation $\Kct (\tilde{\T}_x, \tilde{\T}_y)$ means that we can consider $\tilde{\K}$ as a finite matrix $\Kct: \mathbb{C}^{B_1} \to \mathbb{C}^{W_1}$, with coefficients in the skew Laurent polynomials $\mathbb{C}[q,q^{-1},\tilde{T}_x,\tilde{T}_x^{-1}, \tilde{T}_y,\tilde{T}_y^{-1}]$. For example, this matrix presentation for Kasteleyn operator of the network drawn in Fig.~\ref{fig:todanetwork} is
\begin{equation}
\label{eq:kastexample}
\Kct = 
\left(\begin{array}{cc}
w_1 + w_3 \tilde{\T}_y^{-1} & - w_6 - w_8 \tilde{\T}_x \\
w_4 + w_2 \tilde{\T}_x^{-1} & w_7 + w_5 \tilde{\T}_y \\
\end{array}\right).
\end{equation}

The space $\mathbb{C}^{|\mathbb{Z}^2|}$ as a representation of the algebra of $q$-difference operators by $\tilde{\T}_x$ and $\tilde{\T}_y$ is largely reducible. The degeneracy can be lifted utilizing the algebra of  $q^{-1}$-difference operators, represented by ``dual magnetic translations''
\begin{equation}
\tilde{\T}^{\vee}_x = \sum_{(i,j)\in\mathbb{Z}^2} q^{-\frac{1}{2}j} \, \E_{i-1,i} \otimes \E_{j,j} , ~~~
\tilde{\T}^{\vee}_y = \sum_{(i,j)\in\mathbb{Z}^2} q^{\frac{1}{2}i} \, \E_{i,i} \otimes \E_{j-1,j}, ~~~
\tilde{\T}^{\vee}_y \tilde{\T}^{\vee}_x = q^{-1} \tilde{\T}^{\vee}_x \tilde{\T}^{\vee}_y.
\end{equation}
which commute with the former
\begin{equation}
[\tilde{\T}_{s}, \tilde{\T}^{\vee}_{s'}] = 0,
~~~
s,s'=x,y.
\end{equation}
Therefore any operator, which is a skew Lauren polynomial $\tilde{\Q} = \tilde{\Q}(\tilde{\T}^{\vee}_{x}, \tilde{\T}^{\vee}_{y})$ in $\tilde{\T}_x^{\vee},\tilde{\T}_y^{\vee}$, commutes with $\tilde{\K}$ in the sense that
\begin{equation}
\left(\Id_{\mathbb{C}^{|W_1|}} \otimes \tilde{\Q} \right) \cdot \tilde{\K} = \tilde{\K} \cdot \left(\Id_{\mathbb{C}^{|B_1|}} \otimes \tilde{\Q}\right).
\end{equation}
The form (\ref{eq:kastblock}) of Kasteleyn operator survives under gauge transformations constant inside of fundamental cells, the universal condition determining operators of dual translations is
\begin{equation}
\tilde{\T}^{\vee}_{x} \tilde{\T}_{x} = q^{-\hat{y}} \, ,~~~ 
\tilde{\T}^{\vee}_{y} \tilde{\T}_{y} = q^{\hat{x}} \, ,~~~
q^{\hat{x}}  = \sum_{(i,j)\in\mathbb{Z}^2} q^i \, \E_{i,i} \otimes \E_{j,j}, ~~~
q^{\hat{y}}  = \sum_{(i,j)\in\mathbb{Z}^2} q^j \, \E_{i,i} \otimes \E_{j,j}.
\end{equation}
The operator $\tilde{\Q}$ is hypostasis of eponymous Laurent polynomial from \cite{KO05}, which was shown there to label possible limit shapes of dimer model. In the next section we will show that the complex Burgers equation controlling limits shapes in \cite{KO05} is simply the WKB approximation in $q\to 1$ limit to the spectral problem for the Kasteleyn operator (\ref{eq:kastblock}).

\section{Seiberg-Witten integrability in WKB approximation}
\label{s:WKB}
In this Section we look at the ``melting'' $q\to 1$ limit of vanishing flux for dimer model. The usual arguments of quantum mechanical quasi-classics are applicable to Kasteleyn operator (\ref{eq:kastexample}) in this limit. The main result of this Section is that the free energy (\ref{eq:prepdef}), which is a regularized volume under the ``limit shape'' (\ref{eq:partdim}), satisfies Seiberg-Witten equations (\ref{eq:SWeq}). We will use only the example (\ref{eq:kastexample}) throughout the Section, but all arguments of it can be generalized in a straightforward way.
\subsection{Quasiclassics of vanishing flux at $q \to 1$ and height function of limit shape}
\label{ss:height}
The main observable in dimer models is ``height'' function, which counts portions of dimers oriented ``horizontally'' and ``vertically'' in average configuration. Its meaning becomes more clear, once the configurations of dimer model are interpreted as stepped surfaces.

Let's choose some reference configuration $D_0$ as in (\ref{eq:partfnorm}). As for any $D\in \Dconf(\Gr)$ holds $\partial D = W - B$, the difference $D-D_0$ is a collection of closed and non-intersecting (having no common vertices) cycles on plane, which we interpret as boundaries of ``steps''. The orientation of cycle determines whether its step is upward or downward. Assuming each step to be of heights $1$, the difference of heights between the pair of faces $f_1, f_2$ of $\Gr$ is $\langle p^*_{f_2,f_1}, D-D_0\rangle$, where $p^*_{f_2,f_1}$ is any path on the dual graph $\Gr^*$ connecting $f_1$ and $f_2$ and $\langle\, , \rangle$ is an intersection pairing. Since $\partial(D-D_0) = 0$, the heights difference is independent on choosing of path $p^*_{f_2,f_1}$ for planar $\Gr$. The averaged height function $h: ~ F \times F \to \mathbb{R}$ computes the mean difference of heights over the ensemble of stepped surfaces
\begin{equation}
\label{eq:heightDiff}
h_{f_2,f_1}(\Gr,w;D_0) = h_{f_2,f_1} = \dfrac{1}{\Z(\Gr,w;D_0)}\, \sum_{D\in \Dconf(\Gr)} \langle p^*_{f_2,f_1}, D-D_0\rangle w_{D-D_0}.
\end{equation}
It is clear from this definition, that the fugacity $\db$ in $q=e^{-\db}$ controls the ``volume'' under the stepped surface made out of these loops, since each loop $l = \partial B$ contributes to the statistical weight of configuration in partition function by $\sim e^{-\db \cdot \text{Area}(B)}$. The infinite volume limit corresponds to $\db \to 0$, and the problem of finding the height function and its fluctuations in this limit is called the limit shape problem. 

Due to free-fermionic nature of the model, all correlating functions of any local observables in it can be computed by bare knowledge of two-point Green function $\G$, defined by the equations\footnote{The equation $[\tilde{\Q}, \G] = 0$ has not-clear-yet physical nature, but should be related to the control over boundary conditions of the model, and the exact Green functions from \cite{OR01, OR05} satisfy it.}
\begin{equation}
\label{eq:GreenSchematic}
\tilde{\K} \cdot \G = \Id, ~~~ [\tilde{\Q}, \G] = 0.
\end{equation}
The problem (\ref{eq:GreenSchematic}) for generic $q$ is fully solved only for hexagonal lattices with various boundary conditions using free fermionic vertex operators in \cite{OR01, OR05}. The knowledge of the solution of (\ref{eq:GreenSchematic}) in few leading orders in $\db$ at $\db\to 0$ limit is enough for any purposes of the limit shape problem, but this is still a cumbersome problem. However, the information about height function itself can be heuristically extracted from the structure of $\Ker \tilde{\K} \cap \Ker \tilde{\Q}$, which is the solution of the simpler problem 
\begin{equation}
\label{eq:zeromode}
\tilde{\K} \psi = 0, ~~~ \tilde{\Q} \psi = 0.
\end{equation}
In coordinates $x = \db i, ~ y = \db j$, considered as continuous coordinates on $\mathbb{R}^2$, these equations become
\begin{equation}
\label{eq:greenfunc}
\left\{ \begin{array}{ll}
\sum_{b\in B_1} (\Kct)_{v,b} \left( e^{\frac{1}{2} y - \varepsilon \partial_x}, e^{-\frac{1}{2} x - \varepsilon \partial_y} \right) \psi_{b}(x,y) = 0
\\
\tilde{\Q}\left( e^{\frac{1}{2} y + \varepsilon \partial_x}, e^{-\frac{1}{2} x + \varepsilon \partial_y} \right) \psi_{b}(x,y) = 0
\end{array} \right.
, ~~~ b \in B_1,~ v\in W_1.
\end{equation}
They can be solved order-by-order in $\db$ using standard quasi-classical anzaets for wave-function
\begin{equation}
\label{eq:wavefunc}
\psi_b(x,y) = \exp \left(\frac{\ii}{\db}S^{(0)}_{b}(x,y)+S^{(1)}_b+... ~ \right), ~ b \in B_1.
\end{equation}
In the leading orders $e^{\frac{1}{\db}\#}$ and $\db^{0}$ the consistency conditions for the equations (\ref{eq:zeromode}) become
\begin{equation}
\label{eq:WKBzeroCons}
\left\{
\begin{array}{ll}
P(e^{z},e^{w}) \equiv \det \Kc(e^{z},e^{w}) = 0 \\
Q(e^{\zc},e^{\wc}) = 0 \\
\sum\limits_{b \in B_1} (\Kc)_{v,b} \left( e^z , e^w \right) e^{S_b^{(1)}} = 0
\end{array}
\right. ,
\end{equation}
where $\Kc = \Kct|_{\db=0}$, $Q = \tilde{\Q}|_{\db=0}$ and
\begin{equation}
\label{eq:quasiclhol}
z=\frac{1}{2} y - \ii \partial_x S^{(0)}(x, y), ~
w = -\frac{1}{2} x - \ii \partial_y S^{(0)}(x, y), ~
\zc= \frac{1}{2} y + \ii \partial_x S^{(0)}(x, y), ~
\wc = -\frac{1}{2} x + \ii \partial_y S^{(0)}(x, y).
\end{equation}
Commutativity of $\tilde{\K}$ and $\tilde{\Q}$ implies in the quasiclassical limit that the differential
\begin{equation}
\label{eq:actionquasicl}
dS^{(0)} = \partial_x S^{(0)} dx + \partial_y S^{(0)} dy = \dfrac{\ii}{2}(z dw - w dz) - \dfrac{\ii}{2}(\zc d \wc - \wc d\zc) + \dfrac{\ii}{2}d(\wc z - \zc w)
\end{equation}
is closed, so the quasiclassical action $S^{(0)} = S^{(0)}(x,y)$ can be defined by its integration from. In the simplest case when $Q=P$, the conditions (\ref{eq:WKBzeroCons}) and (\ref{eq:quasiclhol}) can be solved by $\zc=\bar{z},~\wc=\bar{w}$ and one can simplify (\ref{eq:actionquasicl}) to
\begin{equation}
\label{eq:actionPeqQ}
S^{(0)}_{Q=P}(x,y) = \Im \left( \int^{z(x,y)} (w dz - z dw) + \bar{z} w \right) = -2 \cdot \Im \left( \int^{z(x,y)} z dw \right) + 2 \cdot \Re(z) \Im(w),
\end{equation}
which up to exact terms is $(-2)$ times an imaginary part of integral of the meromorphic differential $zdw$, called Seiberg-Witten differential, over the complex curve
\begin{equation}
\curve_P = \{ P(e^z,e^w)=0 \subset (\mathbb{C}^*)^2 \}.
\end{equation}

To compute the height function, let's assume now that the local behaviour of model with flux in $\db \to 0$ limit mimics those of the ``homogeneous'' model of zero flux on the torus. For homogenous model the height function can be easily computed using an expression for free energy density \cite{KOS03}
\begin{equation}
\label{eq:rondef}
\Ron(B_x, B_y) = \int\limits_{0}^{2\pi} \int\limits_{0}^{2\pi} \dfrac{d\theta d\phi}{(2\pi)^2} \log P(e^{B_x + \ii \theta}, e^{B_y + \ii \phi}),
\end{equation}
since the average number of ``horizontal'' and ``vertical'' dimers are dual to the ``twist'' parameters $(B_x, B_y)$
\begin{equation}
\label{eq:hpartdef}
\left\{
\begin{array}{l}
h(x+\db,y) - h(x,y) \simeq - \partial_{B_y} \Ron = \frac{\theta_*}{\pi} \\
h(x,y+\db) - h(x,y) \simeq \partial_{B_x} \Ron = \frac{\phi_*}{\pi}
\end{array}
\right. ,
~~~ \text{where} ~~~
P(e^{B_x + \ii \theta_*}, e^{B_y + \ii \phi_*}) = 0.
\end{equation}
At the same time, the zero-mode of homogeneous model is
\begin{equation}
\label{eq:wavefuncunif}
\psi_{\alpha, (a,b)} = e^{\ii (a \theta_* + b \phi_*)} \xi_\alpha, ~~~ \alpha \in B_1, ~ (a,b)\in \mathbb{Z}^2,
~~~ \text{where} ~~~
(\K_1)(e^{B_x + \ii \theta_*}, e^{B_y + \ii \phi_*}) \cdot \xi = 0.
\end{equation}
Applying in (\ref{eq:wavefuncunif}) coordinates $a = x/\db, ~ b = y/\db$ and comparing it with (\ref{eq:wavefunc}), one can guess the height function of the model with flux in $\db \to 0$ limit to be
\begin{equation}
\label{eq:heightfunc}
h(x,y) = \int \left( \partial_x h \, dx + \partial_y h \, dy \right) \simeq \frac{S^{(0)}(x,y)}{\pi \db}.
\end{equation}
The WKB quantization condition coming from single-valuedness of wave-function becomes also the natural condition for height difference between frozen regions of the model \cite{KO05} to be integral.

In the case of $Q=P$ comparing formulas (\ref{eq:WKBzeroCons}), (\ref{eq:quasiclhol}) with (\ref{eq:hpartdef}), one can deduce
\begin{equation}
h_{Q=P}(x,y) = - \dfrac{2}{\db} \Ron\left(\frac{y}{2},-\frac{x}{2}\right).
\end{equation}

In \cite{KO05} similar results were obtained, but the logic (and notations) were different. Pair of equations (\ref{eq:WKBzeroCons}) appeared there as a solution of variational problem, optimizing the total surface tension\footnote{The surface tension density is a Legandre dual to the free energy density $\Ron$. It computes the energy of the region with the known slope $(\partial_x h, \partial_y h)$ in opposite to $\Ron$, which computes energy of the region with fugacities $(B_x, B_y)$.} to be minimal. The Euler-Lagrange equation of this problem results to equations
\begin{equation}
\partial_y z  - \partial_x w = 1, ~~~ P(e^z,e^w) = 0,
\end{equation}
called complex Burgers equation. The function $\Q$ appears then as a free function, parametrizing the space of solutions of this equation, and controlling the boundary conditions for solutions. So the equation, which in our setup is a consistency condition supporting Hamilton-Jacobi equation, appears also to be the stationary-action principle for $2d$ field theory. Expression for height function similar to (\ref{eq:heightfunc}) was also derived in \cite{KO05}.

\subsection{Free energy density is Seiberg-Witten prepotential.}
\label{ss:SWpot}
The WKB arguments can be also applied to computation of partition function in $\db\to 0$ limit. The usual heuristics
\begin{equation}
\mathrm{Tr}[A(T_x,T_y)] ~ \to ~ \iint \dfrac{dx dy}{\db^2} \iint \dfrac{d\theta d\phi}{(2\pi)^2} A(e^{\frac{y}{2} + \ii \theta}, e^{-\frac{x}{2} + \ii \phi}) \text{ ~ as ~ } \db \to 0
\end{equation}
gives the integral formula for the partition function of the model
\begin{equation}
\label{eq:partdim}
\Z = \det \tilde{\K} = e^{\mathrm{tr} \log \tilde{\K}} \propto \exp\left( \dfrac{1}{\db^2} \iint dxdy \, \Ron\left(\dfrac{y}{2}, - \dfrac{x}{2} \right) \right) = q^{\Vol (P, P)},
~
\Vol(P,P) = \frac{1}{2} \iint \dfrac{dx dy}{\db^2} h_{Q=P}(x,y).
\end{equation}
The proportionality of the free energy of the model to the volume\footnote{Up to $1/2$, whose appearance in the definition of $\Vol$ is unclear.} under the limit shape is a natural thing: in the leading order, the partition function is dominated by single configuration, and the free energy determined by it is proportional to the sum of areas of all contours which this configuration contains (which is basically volume). It is diverging, and proper regularization of determinant in (\ref{eq:partdim}) and extension of the formula to the case $Q \neq P$ requires careful consideration of the boundary conditions for the model and role of $Q$. We will instead define some regularization of $\Vol$ guided by its properties and natural equation satisfied by it. In order to to this we need first to make a closer look to the properties of spectral curve $P(e^{z},e^{w}) = 0$ and function $\Ron$.\\

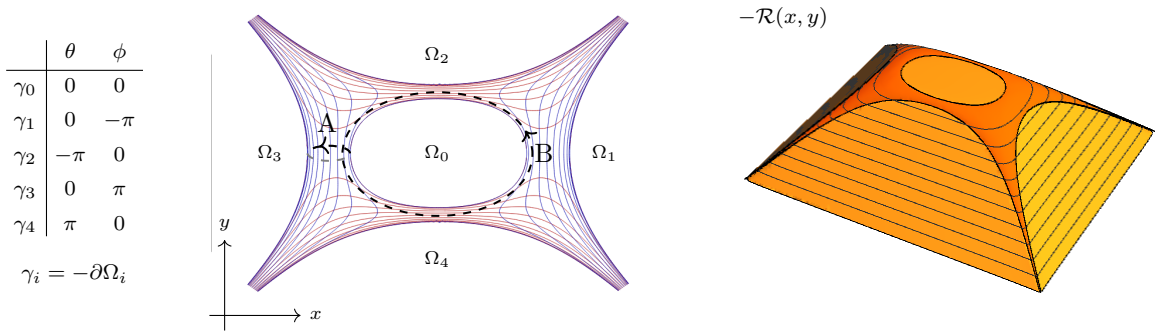
\begin{figure}[!h]
\begin{center}
\begin{tikzpicture}

\node at (0,-1.8) {\footnotesize $\gamma_i = - \partial \Dom_i$};

\matrix [
	matrix of nodes,
	row sep=-\pgflinewidth,
	font=\footnotesize,
	]
{
	& $\theta$ & $\phi$ \\
	$\gamma_0$   & $0$ & $0$ \\
	$\gamma_1$   & $0$ & $-\pi$ \\
	$\gamma_2$   & $-\pi$ & $0$ \\
	$\gamma_3$   & $0$ & $\pi$ \\

	$\gamma_4$   & $\pi$ & $0$ \\
};

\draw (-0.9,0.9)--(0.9,0.9);
\draw (-0.37,1.3)--(-0.37,-1.3);
\draw[white] (0.5,-2.7)--(0,-2.7);

\end{tikzpicture}
\scalebox{1}{
\begin{tikzpicture}

\draw [white, path picture={
		\node at (0,0) {\includegraphics[width=.35\textwidth]{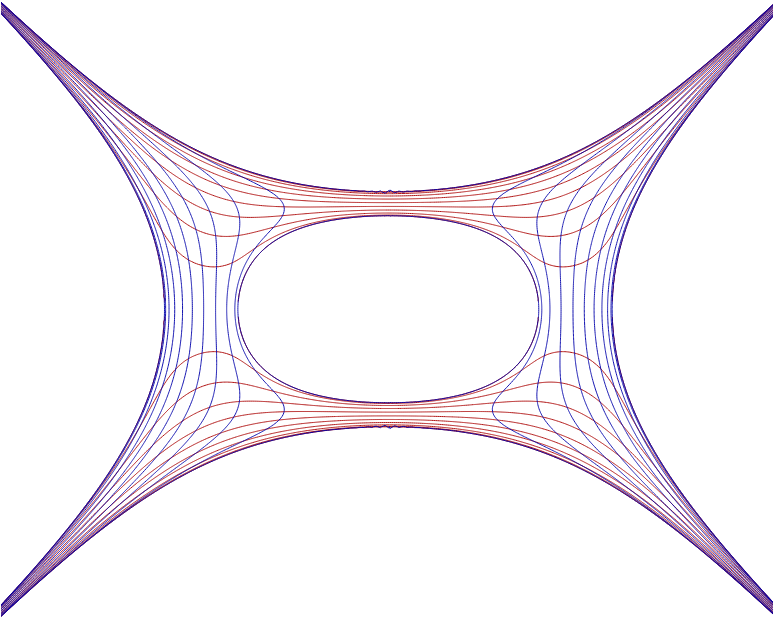}};
	};] (0,0) ellipse (3.5 and 2.5);

\draw[line width=0.75, dashed, gray] (-1.15,0) arc(0:360:0.28 and 0.1);

\draw[
	decoration={markings, mark=at position 0.7 with {\arrow{>}}},
	postaction={decorate},
	line width=0.75, dashed
	] (-1.15,0) arc(0:180:0.28 and 0.1);

\draw[
	decoration={markings, mark=at position 0.05 with {\arrow{>}}},
	postaction={decorate},
	line width=0.75, dashed
	] (0,-0.01) ellipse(1.25 and 0.82);

\node at (-1.45,0.4) {$\mathrm{A}$};
\node at (1.4,0) {$\mathrm{B}$};

\draw [->] (-3,-2.15)--(-1.8,-2.15);
\draw [->] (-2.8,-2.35)--(-2.8,-1.15);
\node at (-1.6,-2.15) {\scriptsize $x$};
\node at (-2.8,-0.95) {\scriptsize $y$};

\node at (0,0) {\scriptsize $\Dom_0$};
\node at (2.2,0) {\scriptsize $\Dom_1$};
\node at (0,1.3) {\scriptsize $\Dom_2$};
\node at (-2.2,0) {\scriptsize $\Dom_3$};
\node at (0,-1.4) {\scriptsize $\Dom_4$};

\end{tikzpicture}}
\hspace*{0.1cm}
\begin{tikzpicture}
\node at (0,0) {\includegraphics[width=.32\textwidth]{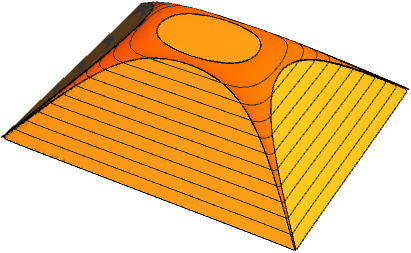}};
\node at (-2.2,2) {\footnotesize $-\Ron(x,y)$};
\draw[white] (0,-2.3)--(1,-2.3);
\end{tikzpicture}
\end{center}
\caption{Left: Amoeba $\Am(P)$ of the curve $P(e^{x+\ii\theta},e^{y+\ii\phi}) = 0$. Red lines are for $\theta=\mathrm{const}$, blue are for $\phi=\mathrm{const}$, their values are taken for one of two sheets of $\curve_P$ over $\Am(P)$. Boundaries $\gamma_i$ of ovals $\Dom_i$ are oriented counter-clockwise along $\partial \Am$. The projections of $A-$ and $B-$ cycles are drawn by dashed lines. Right: minus Ronkin function $-\Ron(x,y)$ for the same $P$.}
\label{fig:amoebaToda}
\end{figure}

For the lattice drawn on Fig. \ref{fig:todanetwork}, the Laurent polynomial $P$ computed using (\ref{eq:kastexample}) is
\begin{equation}
\label{eq:curveToda}
P(\lambda,\mu) = \det \Kc(\lambda,\mu) = \dfrac{w_2 w_6}{\lambda} + w_4 w_8 \lambda + w_1 w_5 \mu + \dfrac{w_3 w_7}{\mu} + \left( w_3 w_5 + w_2 w_8 + w_1 w_7 + w_4 w_6 \right).
\end{equation}
For the purposes of this Section the rescalings $P(\lambda,\mu) \mapsto A \cdot P(B\lambda, C \mu)$ are immaterial, so we will be using here $P$ in the equivalent form
\begin{equation}
\label{eq:LaurentPoly}
P(\lambda,\mu) = \lambda + \dfrac{Z}{\lambda} + \mu + \dfrac{1}{\mu} - U,
\end{equation}
\begin{equation}
Z = x_1 x_3, ~~~
-U = \sqrt{x_1 x_4}+\frac{1}{\sqrt{x_1 x_4}} + \sqrt{Z} \left(\sqrt{x_3 x_4} + \frac{1}{\sqrt{x_3 x_4}}\right),
\end{equation}
where $x_i$ are face variables labelled following Fig.~\ref{fig:todanetwork}, left. Curves $\curve_P$ appearing in planar dimer models are Harnak \cite{KO03}, which means that the logarithmic projection $(\lambda,\mu) \mapsto (\log |\lambda|,\log |\mu|)$ of spectral curve $\curve_P$ to $\mathbb{R}^2$ is $2$ to $1$ mapping in the interior of amoeba\footnote{Starting from here and until the end of this Section we use coordinates $(x,y)$ differently compared to the usage above.}
\begin{equation}
\Am(P) = \{ (x, y)\in \mathbb{R}^2 ~ | ~ \exists ~ (\theta, \phi) \in \mathbb{R}^2:  P(e^{x+\ii \theta}, e^{y+\ii\phi}) = 0 \},
\end{equation} 
and $1$ to $1$ at its boundary. The inverse is also true: any Harnak curve in $\mathbb{C}^* \times \mathbb{C}^*$ can be obtained from some planar dimer model. For Laurent polynomial (\ref{eq:curveToda}) the curve is Harnak if $Z\in \mathbb{R}_{\geq 0}$, $U\geq U_0 = 2(\sqrt{Z}+1)$ which is satisfied because of $x_i\in \mathbb{R}_{\geq 0}$, following from positivity of edge weights. The corresponding amoeba is drawn on Fig.~\ref{fig:amoebaToda}, left.

Complement of amoeba of Harnak curve consists of disjoint regions $\mathbb{R}^2 \backslash \Am(P) = \cup_i \Dom_i$, which are bounded and unbounded ovals. Their combinatorics of ovals is captured by Newton polygon $N_P$ of polynomial $P$ - the convex hull of such $(i,j) \in \mathbb{Z}^2$, that $\lambda^i \mu^j$ is contained in $P(\lambda,\mu)$ with non-zero coefficient. Bounded ovals correspond to integral internal points of $N_P$, unbounded ovals to integral boundary points, so the amoeba can be contracted to the graph, dual to some triangulation of Newton polygon. The function $\Ron$, called Ronkin function of $P$ in mathematical literature, in case of Harnak $P$ is concave function on $\mathbb{R}^2$, linear of slope $(i,j)$ on oval corresponding to point $(i,j)$ of Newton polygon, and interpolating slopes of ovals in the interior of amoeba, as shown on Fig.~\ref{fig:amoebaToda}, right.

Since the ovals have to be invariant under the complex involution $(\lambda,\mu) \mapsto (\bar{\lambda},\bar{\mu})$, functions $\theta(x,y)$ and $\phi(x,y)$ can take only $\pi\mathbb{Z}$ values there. The parametrization of $\curve_P$ by $(z,w)$ is uniquely determined by the condition, that the single-valued smooth functions $\theta(x,y)$, $\phi(x,y)$ in the interior of $\Am$ are such solution of
\begin{equation}
z = x + i \theta(x,y),~~~
w = y + i \phi(x,y):~~~
P(e^{z},e^{w}) = 0,
\end{equation}
that $\theta = \phi = 0$ at $\gamma_0$ and $\phi(x,y)$ is increasing along the short paths from $\gamma_0$ to $\gamma_3$. We call part of $\curve_P$ parametrized by this $(z,w)$ to be upper sheet, and those, which is complex conjugated, to be lower. Both $\theta$, $\phi$ considered as a functions on $\curve_P$ are single valued in the interior of $\Am$ and on $\gamma_0$, however they can have jumps at other $\gamma_i$.

Now we can define the regularization of free energy in (\ref{eq:partdim}) by
\begin{equation}
\label{eq:prepdef}
\prep(U) = \tilde{\prep}(U) - \tilde{\prep}(U_0), ~~~
\tilde{\prep}(U) = \dfrac{\ii}{\pi} \left(\iint\limits_{\mathbb{R}^2} \Ron(x,y) dx dy - \left(\int\limits_{\gamma_1}-\int\limits_{\gamma_3}\right) \dfrac{x^2 dy}{8} - \left(\int\limits_{\gamma_4}-\int\limits_{\gamma_2}\right) \dfrac{y^2 dx}{8} \right).
\end{equation}
It is finite, since at large $x,y$ graphs or Ronkin functions for $P$ with the same values of $Z$ but different $U$ are exponentially close. The overall normalization and presence of boundary terms is justified by the following Claim, which is natural due to the reasons explained in Introduction:
\paragraph{Claim.} The prepotential $\prep$ defined in (\ref{eq:prepdef}) satisfies Seiberg-Witten equation
\begin{equation}
\label{eq:SWeq}
\dfrac{\partial \prep}{\partial a} = a_D,
~~~~~
a = \oint_A z \dfrac{dw}{2\pi \ii},
~~~~~
a_D = \oint_B z \dfrac{dw}{2\pi \ii},
\end{equation} 
where $A$ and $B=-\gamma_0$ are simple cycles on curve, which intersect with $A\cap B = 1$, as shown on Fig.~\ref{fig:amoebaToda}, and orientation of $A$-cycle is such, that it is directed from $\gamma_0$ to $\gamma_3$ when goes along the upper sheet of $\curve_P$.

\paragraph{Proof.}{Firstly, note that $a = a(U)$ is analytic function at a generic point, so (\ref{eq:SWeq}) is equivalent to
\begin{equation}
\label{eq:SWbyU}
\dfrac{\partial \prep}{\partial U} = a_D  \dfrac{\partial a}{\partial U},
\end{equation}
and that since $\Ron(x,y; U)-\Ron(x,y; U_0)$ is exponentially small at infinity, we can interchange integration and differentiation
\begin{equation}
\dfrac{\partial}{\partial U}\iint\limits_{\mathbb{R}^2} (\Ron(x,y; U)-\Ron(x,y; U_0)) ~ \dfrac{dx\wedge dy}{2\pi \ii}
=
\iint\limits_{\mathbb{R}^2} \dfrac{\partial \Ron(x,y)}{\partial U} ~ \dfrac{dx\wedge dy}{2\pi \ii}.
\end{equation}
Decompose $\mathbb{R}^2 = \Dom_0 \cup \Am \cup \Dom_1 \cup \Dom_2 \cup \Dom_3 \cup \Dom_4$, and consider integrals over the regions separately. For any of $\Dom_i$ or $\Am$, their shapes depend on $U$, so change of the order of differentiation and integration over any single of them would change integral by additional contact term.
\begin{itemize}

\item Let $(x,y)\in \Dom_0$, then
\begin{equation}
\label{eq:SWderivD0}
\dfrac{\partial \Ron(x,y)}{\partial U} = 
\int\limits_{0}^{2\pi} \dfrac{d \phi}{2\pi} \oint\limits_{|z|=x} \dfrac{dz}{2\pi \ii} ~ \dfrac{\partial_U P (e^{z},e^{y+\ii \phi})}{P (e^{z},e^{y+\ii \phi})} = 
\int\limits_{0}^{2\pi} \dfrac{d \phi}{2\pi} ~  \dfrac{\partial_U P (e^{z_*},e^{y+\ii \phi})}{\partial_z P (e^{z_*},e^{y+\ii \phi})} = - \oint_A \dfrac{\partial z(w)}{\partial{U}} \dfrac{dw}{2\pi \ii} = - \dfrac{\partial a(U)}{\partial U} 
\end{equation}
where the contour of integration is deformed first from $\Re z=x$ to $\Re z = -\infty$, keeping $\Re w = y$, and picking pole at $z_*$, such that $P(e^{z_*},e^{y+\ii \phi})=0$, see Fig.~\ref{fig:contour}. Then the remaining integration over $d\phi$ becomes integral of $-\ii dw$ over $A$-cycle, and we use that $0 = dP/dU = \partial_U P + \partial_z P \partial_U z$, assuming that $z=z(U,w)$ \footnote{These two steps are equivalent to deformation of $2d$ contour and picking Poincar\'e residue of $\frac{dz\wedge dw}{P}$ at $P=0$}. As $\partial_U \Ron(x,y)$ does not depends on $(x,y)\in \Omega_0$, it remains to compute
\begin{equation}
\iint\limits_{\Dom_0} \dfrac{dx\wedge dy}{2\pi \ii} =
\oint\limits_{\partial \Dom_0} \dfrac{x dy}{2\pi \ii} =
\dfrac{1}{2\pi \ii} \oint\limits_{-\gamma_0} (z dw - \ii(\theta dy + x d\phi) + \theta d \phi) =
\oint_{B} z \dfrac{dw}{2\pi \ii}  = a_D(U)
\end{equation}
where we used that $\theta=\phi=0$ at $\gamma_0$.

\begin{figure}[!h]
\begin{center}
\begin{tikzpicture}
\draw[white] (0,0)--(0.3,0);
\end{tikzpicture}
\scalebox{1}{
\begin{tikzpicture}
\node[inner sep=0pt] at (0,0) {\includegraphics[width=.5\textwidth]{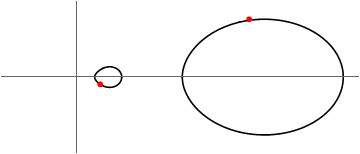}};
\draw[
	decoration={markings, mark=at position 0.625 with {\arrow{>}}},
	postaction={decorate},
	dashed, thick
]	(-2.4,0) circle (1.5);

\node at (-1, -1.6) {\footnotesize $\mathrm{Re}\, z = x$};
\node at (-1.7, -0.5) {\footnotesize $z_*(\phi)$};

\node[draw=white, circle, fill=blue!10, ultra thin] at (-4, 1.6) {$e^{z}$};
\draw[white] (0,-2.1)--(1,-2.1);
\end{tikzpicture}
}
\end{center}
\caption{Slice of the curve $\curve_P$ by $y=\mathrm{const}$ section, plotted in $e^{z}$ coordinate, shown by ovals. The $y$ is such that the $y=\mathrm{const}$ line crosses a hole of amoeba. Red dots are points with the same $\phi$. Dashed circle is $dz$ integration contour in (\ref{eq:SWderivD0}), which has to be contracted to zero.}
\label{fig:contour}
\end{figure}
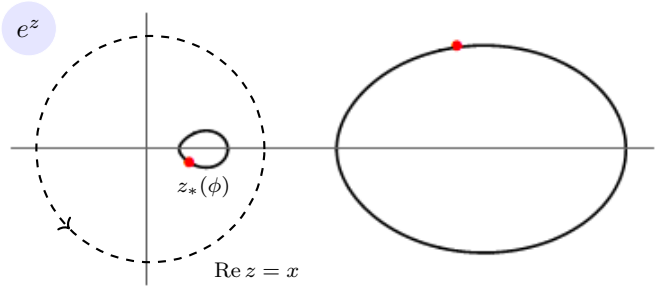

\item Regions $\Dom_1, \Dom_2, \Dom_3, \Dom_4$ do not contribute to integral, as we can deform integration contour there to $\Re z \to +\infty$, $\Re w \to +\infty$, $\Re z \to -\infty$, $\Re w \to -\infty$ respectively, where integrand is exponentially suppressed, without picking any poles.

\item For any $(x,y)\in \Am$ we can shift integration contour to $x \to -\infty$, along any sequence of straight segments of rational slope. The poles are picked as in (\ref{eq:SWderivD0}), because of $\mathrm{SL}(2,\mathbb{Z})$ invariance of integration measure
\begin{equation}
\label{eq:SL2Zinv}
-\dfrac{\partial z(w)}{\partial{U}} \dfrac{dw}{2\pi \ii} =
\dfrac{\partial_U P}{\partial_z P} \dfrac{dw}{2\pi \ii} =
\dfrac{\partial_U P}{d \partial_{\tilde{z}} P - c \partial_{\tilde{w}} P } \left(d  + c \dfrac{\partial\tilde{z}}{\partial\tilde{w}} \right) \dfrac{d \tilde{w}}{2\pi \ii} =
-\dfrac{\partial \tilde{z}(\tilde{w})}{\partial{U}} \dfrac{d\tilde{w}}{2\pi \ii},
\end{equation}
where $z = a \tilde{z} + b \tilde{w}$, $w = c \tilde{z} + d \tilde{w}$, with $a,b,c,d\in \mathbb{Z}$, $ad-bc=1$. As the integrand is a holomorphic form, the integration contour might be deformed to any convenient smooth contour which goes from $w = w(x,y)$ to $\gamma_3$, and then to $\bar{w}$, on another sheet. Using that inside of $\Am$ we can present area element $dx \wedge dy$ as
\begin{equation}
dx\wedge dy = \frac{1}{4} \left(dz\wedge d\bar{w} + d\bar{z} \wedge dw \right),
\end{equation}
we apply integration by parts, to get
$$
- \iint_{\Am} \left( \int_{w}^{\bar{w}} \dfrac{\partial z(w)}{\partial U} \dfrac{dw}{2\pi \ii} \right) \dfrac{dz\wedge d\bar{w} + d\bar{z} \wedge dw}{8\pi \ii} = 
$$
\begin{equation}
=
\int_{\partial \Am} \left(\int_{w}^{\bar{w}} \dfrac{\partial z(w)}{\partial U} \dfrac{dw}{2\pi \ii} \right) \dfrac{\bar{w} dz + w d\bar{z}}{8\pi \ii} +
\iint_{\Am} \left( \bar{w}\dfrac{\partial \bar{z}(\bar{w})}{\partial U} \dfrac{dz\wedge d\bar{w}}{(4\pi \ii)^2 } - w \dfrac{\partial z(w)}{\partial U} \dfrac{d\bar{z}\wedge dw}{(4\pi \ii)^2 } \right)
=
\end{equation}
$$
= \int_{\partial \Am} \left(\int_{w}^{\bar{w}} \dfrac{\partial z(w)}{\partial U} \dfrac{dw}{2\pi \ii} \right) \dfrac{\bar{w} dz + w d\bar{z}}{8\pi \ii} +
\int_{\partial \Am} \left( \dfrac{\partial \bar{z}(\bar{w})}{\partial U} \dfrac{z\bar{w} d\bar{w}}{(4\pi \ii)^2 } -  \dfrac{\partial z(w)}{\partial U} \dfrac{\bar{z}w dw}{(4\pi \ii)^2 } \right).
$$
Using that the contours in $\int_{w}^{\bar{w}} (\partial z/\partial U) dw$ are now closed (since $w=\bar{w}$ at $\partial \Am$), and some of them can be contracted to points at infinity, where $\partial z(w)/\partial U$ is exponentially suppressed, the first integral reduces to
\begin{equation}
\sum_{i=0}^{4} \int\limits_{\gamma_i} \left(\int_{w}^{\bar{w}} \dfrac{\partial z(w)}{\partial U} \dfrac{dw}{2\pi \ii} \right) \dfrac{\bar{w} dz + w d\bar{z}}{8\pi \ii} =
\dfrac{\partial a}{\partial U} \int\limits_{\gamma_0} \dfrac{y dx}{4\pi \ii} = \dfrac{a_D}{2} \dfrac{\partial a}{\partial U}.
\end{equation}
Using also the values of $\theta,\phi \in \pi \mathbb{Z}$ on $\gamma_i$ at upper sheet of $\curve_P$, which are indicated on Fig.~\ref{fig:amoebaToda}, and $\mathrm{SL}(2,\mathbb{Z})$ invariance (\ref{eq:SL2Zinv}), we get for the remaining
\begin{equation}
\sum_{i=0}^{4} \int\limits_{\gamma_i} \left( \dfrac{\partial \bar{z}(\bar{w})}{\partial U} \dfrac{z\bar{w} d\bar{w}}{(4\pi \ii)^2 } -  \dfrac{\partial z(w)}{\partial U} \dfrac{\bar{z}w dw}{(4\pi \ii)^2 } \right) = 
\sum_{i=0}^{4} \int\limits_{\gamma_i} \dfrac{\partial z}{\partial U} \dfrac{(z\bar{w} - \bar{z}w) dw }{(4\pi \ii)^2} = 
\int\limits_{\gamma_1-\gamma_3} \dfrac{\partial x}{\partial U} \dfrac{xdy}{8 \pi \ii} + \int\limits_{\gamma_4-\gamma_2} \dfrac{\partial y}{\partial U} \dfrac{ydx}{8 \pi \ii} .
\end{equation}

\end{itemize}

All contributions brought together give us identity (\ref{eq:SWbyU}). $\hfill \blacksquare$}\\

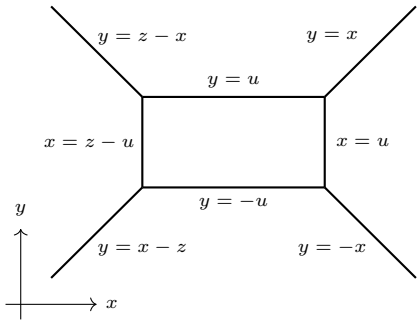
\begin{figure}[!h]
\begin{center}
\begin{tikzpicture}

\draw[thick] (-1.2,-0.6) -- (-1.2,0.6);
\draw[thick] (-1.2,0.6) -- (1.2,0.6);
\draw[thick] (1.2,0.6) -- (1.2,-0.6);
\draw[thick] (1.2,-0.6) -- (-1.2,-0.6);

\draw[thick] (-1.2, -0.6) -- (-2.4, -1.8);
\draw[thick] (1.2, -0.6) -- (2.4, -1.8);
\draw[thick] (-1.2, 0.6) -- (-2.4, 1.8);
\draw[thick] (1.2, 0.6) -- (2.4, 1.8);

\draw [->, thin] (-3,-2.15)--(-1.8,-2.15);
\draw [->, thin] (-2.8,-2.35)--(-2.8,-1.15);
\node at (-1.6,-2.15) {\scriptsize $x$};
\node at (-2.8,-0.9) {\scriptsize $y$};

\node at (0,-0.8) {\scriptsize $y=-u$};
\node at (-1.2,-1.4) {\scriptsize $y=x-z$};
\node at (1.3,-1.4) {\scriptsize $y=-x$};

\node at (0,0.8) {\scriptsize $y=u$};
\node at (-1.2,1.4) {\scriptsize $y=z-x$};
\node at (1.3,1.4) {\scriptsize $y=x$};

\node at (1.7,0) {\scriptsize $x=u$};
\node at (-1.9,0) {\scriptsize $x=z-u$};

\end{tikzpicture}
\end{center}
\caption{Amoeba of the curve $\curve_P$ in tropical limit. Coordinates here are normalized by $\R5$}
\label{fig:amoebaTrop}
\end{figure}

Another interesting limit can be taken now. It is called perturbative or tropical or decompactification in different contexts. In it, the parameters scale as
\begin{equation}
U = e^{\R5 u}, ~~~ Z = e^{\R5 z}, ~~~ \R5 \to +\infty.
\end{equation}
The amoeba shrinks then to its spine, which is a union of intervals as shown on Fig.~\ref{fig:amoebaTrop}, and pre-image of projection $\curve_P \to \Am$ becomes $S^1$ over the internal points of intervals, and pairs of triangles, connecting these circles, over the joints of intervals. The Ronkin function in the leading in $\R5$ order become piecewise linear function of $x,y$, and integrations in (\ref{eq:prepdef}) becomes trivial exercises in computations of polyhedron volumes. Taking $U_0 = 2(\sqrt{Z}+1)$ at which domain $\Dom_0$ shrinks to point, one gets 
\begin{equation}
\label{eq:prepTrop}
\prep \to - \dfrac{\R5^3}{24 \pi \ii} (2 u - z)^2 (4 u + z),
~~~
a = \oint_A z \dfrac{dw}{2\pi \ii} \to \R5 \cdot (z-u),
~~~
a_D = \oint_B z \dfrac{dw}{2\pi \ii} \to \dfrac{\R5^2}{2\pi \ii} 2u(2u-z).
\end{equation}
This completely ``frozen'' by extreme values of parameters configuration will be the starting point in the next Section. However we will ``unfroze'' it in a different way, keeping finite $q$ under extreme values of $x_i$.

\section{Boxcounting in tropical limit}
\label{s:boxes}

In this Section we will show, how the Fourier-transformed topological string amplitude (\ref{eq:ZinstD}) comes combinatorially from the counting of dimers in the running example as on Fig.~\ref{fig:todanetwork}: we identify degrees of freedom corresponding to $0d$ boxes constituting $3d$ Young diagrams, $1d$ boxes constituting $2d$ Young diagrams and $2d$ boxes constituting $1d$ Young diagrams. We also suggest how the properly taken tropical limit for face weights $x_i = e^{\R5 \xi_i + \we{x}_i}, ~ \R5 \to \infty$ might suppress all the other degrees of freedom, but it appears to be inconsistent with the thermodynamic limit.

\subsection{Combinatorics of boxcounting}
\begin{figure}[!h]
\centering
\scalebox{0.65}{
\begin{tikzpicture}

\tikzmath{
	\N=9;\M=9;
	\scale=1;
	\cofset=0.8;
	\L=1;
	\fsize=1.75;
	\lwidth=1mm;
}

\node at (9.1*\L+\L/2, \cofset/2) {\Large $Q_0 = x_1 x_2 x_3 x_4$};
\node at (22.35*\L+\cofset, \cofset/2) {\Large $Q_{1,B} = x_2 x_3 (Q_0)^{N}$};
\node at (9.1*\L+\L/2, \cofset/2-16*\L) {\Large $Q_{1,F} = \dfrac{x_2}{x_1} (Q_0)^{M}$};
\node at (22.35*\L+\cofset, \cofset/2-16*\L) {\Large $Q_{2} = x_2 \left( \frac{x_2}{x_1}\right)^{N} (x_2 x_3)^{M} (Q_0)^{NM}$};

\begin{scope}[scale=\scale]

\clip(4*\L-\cofset,2*\L-\cofset) rectangle (2*\M - 3*\L + \cofset, 2*\N - 3*\L + \cofset);
\tikzmath{
		{
			\filldraw[fill=lime] (5*\L,3*\L) rectangle ++(\L,2*\L);	
			\filldraw[fill=lime] (6*\L,4*\L) rectangle ++(\L,2*\L);		
		};
		{
			\filldraw[fill=lime] (13*\L,3*\L) rectangle ++(\L,2*\L);	
			\filldraw[fill=lime] (12*\L,4*\L) rectangle ++(\L,2*\L);		
		};
		{
			\filldraw[fill=lime] (5*\L,12*\L) rectangle ++(\L,2*\L);	
			\filldraw[fill=lime] (6*\L,11*\L) rectangle ++(\L,2*\L);		
		};
		{
			\filldraw[fill=lime] (13*\L,12*\L) rectangle ++(\L,2*\L);	
			\filldraw[fill=lime] (12*\L,11*\L) rectangle ++(\L,2*\L);		
		};
		for \i in {1,...,2*\M}{
			{			
				\draw (\i*\L,-\L*0.5) -- (\i*\L,\L*2*\N-\L*0.5);
			};
		};
		for \i in {0,...,2*\N-1}{
			{			
				\draw (\L*0.5,\i*\L) -- (\L*2*\M+\L*0.5,\i*\L);
			};
		};
		for \i in {0,...,\M}{
			{
				\draw[color=gray, dashed] (2*\i*\L+0.5*\L,-\L*0.5) -- (2*\i*\L+0.5*\L,\L*2*\N-\L*0.5);
			};
		};
		for \i in {0,...,\N}{
			{			
				\draw[color=gray, dashed] (\L*0.5,2*\i*\L-0.5*\L) -- (\L*2*\M+\L*0.5,2*\i*\L-0.5*\L);
			};
		};
		for \x in {1,...,\M}{
			for \y in {0,...,\N-1}{
				{			
					\draw[fill=white] (2*\L*\x,2*\L*\y) circle[radius=0.05];
				};
			};
		};
		for \x in {0,...,\M-1}{
			for \y in {0,...,\N-1}{
				{			
					\draw[fill=white] (2*\L*\x+\L,2*\L*\y+\L) circle[radius=0.05];
				};
			};
		};
		for \x in {0,...,\M-1}{
			for \y in {0,...,\N-1}{
				{			
					\draw[fill] (2*\L*\x+\L,2*\L*\y) circle[radius=0.05];
				};
			};
		};	
		for \x in {1,...,\M}{
			for \y in {0,...,\N-1}{
				{			
					\draw[fill] (2*\L*\x,2*\L*\y+\L) circle[radius=0.05];
				};
			};
		};	
		for \x in {1,...,\M-1}{
			for \y in {1,...,\N-1}{
				{
					\tikzmath{\la=int(-\x+1);\lb=int(-\x+2);};		
					\node[scale=\fsize] at (2*\L*\x+0.5*\L,2*\L*\y-0.5*\L) {$_1$};
				};
			};
		};
		for \x in {0,...,\M-1}{
			for \y in {1,...,\N-1}{
				{
					\tikzmath{\la=int(\x-1);\lb=int(\x-1);};		
					\node[scale=\fsize] at (2*\L*\x+1.5*\L,2*\L*\y-0.5*\L) {$_2$};
				};
			};
		};
		for \x in {0,...,\M-1}{
			for \y in {0,...,\N-1}{
				{
					\tikzmath{\la=int(-\x+1);\lb=int(-\x+2);};
					\node[scale=\fsize] at (2*\L*\x+1.5*\L,2*\L*\y+\L-0.5*\L) {$_3$};
				};
			};
		};
		for \x in {1,...,\M-1}{
			for \y in {0,...,\N-1}{
				{
					\tikzmath{\la=int(\x-2);\lb=int(\x-2);};		
					\node[scale=\fsize] at (2*\L*\x+0.5*\L,2*\L*\y+\L-0.5*\L) {$_4$};
				};
			};
		};		
		for \y in {0,...,2}{
			for \x in {1,...,\M-2*\y-1}{
				{			
					\draw[color=blue, line width = \lwidth] (2*\L*\x+2*\L*\y,2*\L*\y)--(2*\L*\x+2*\L*\y+\L,2*\L*\y);
				};
			};
		};
		for \y in {0,...,1}{
			for \x in {1,...,\M-2*\y-2}{
				{			
					\draw[color=blue, line width = \lwidth] (2*\L*\x+2*\L*\y+\L,2*\L*\y+\L)--(2*\L*\x+2*\L*\y+2*\L,2*\L*\y+\L);
				};
			};
		};		
		for \y in {0,...,2}{
			for \x in {1,...,\M-2*\y-1}{
				{			
					\draw[color=purple, line width = \lwidth] (2*\L*\x+2*\L*\y,-2*\L*\y+2*\L*\N-\L)--(2*\L*\x+2*\L*\y+\L,-2*\L*\y+2*\L*\N-\L);
				};
			};
		};
		for \y in {0,...,1}{
			for \x in {1,...,\M-2*\y-2}{
				{			
					\draw[color=purple, line width = \lwidth] (2*\L*\x+2*\L*\y+\L,-2*\L*\y+2*\L*\N-2*\L)--(2*\L*\x+2*\L*\y+2*\L,-2*\L*\y+2*\L*\N-2*\L);
				};
			};
		};
		for \x in {1,...,3}{
			for \y in {0,...,\N-2*\x}{
				{			
					\draw[color=olive, line width = \lwidth] (2*\L*\x,2*\L*\y+2*\L*\x)--(2*\L*\x,2*\L*\y+2*\L*\x-\L);
				};
			};
		};
		for \x in {1,...,3}{
			for \y in {-1,...,\N-2*\x}{
				{			
					\draw[color=olive, line width = \lwidth] (2*\L*\x-\L,2*\L*\y+2*\L*\x+\L)--(2*\L*\x-\L,2*\L*\y+2*\L*\x);
				};
			};
		};
		for \x in {1,...,3}{
			for \y in {0,...,\N-2*\x}{
				{			
					\draw[color=orange, line width = \lwidth] (2*\L*\M-2*\L*\x+\L,2*\L*\y+2*\L*\x)--(2*\L*\M-2*\L*\x+\L,2*\L*\y+2*\L*\x-\L);
				};
			};
		};
		for \x in {1,...,3}{
			for \y in {-1,...,\N-2*\x}{
				{			
					\draw[color=orange, line width = \lwidth] (2*\L*\M-2*\L*\x+2*\L,2*\L*\y+2*\L*\x+\L)--(2*\L*\M-2*\L*\x+2*\L,2*\L*\y+2*\L*\x);
				};
			};
		};
		for \x in {2,...,\M-2}{
			for \y in {1,...,\N-5}{
				{			
					\draw[color=violet, line width = \lwidth] (2*\L+\x*\L+3*\L,4*\L+2*\y*\L)--(2*\L+\x*\L+3*\L,4*\L+2*\y*\L-\L);
				};
			};
		};		
};

\end{scope}
\begin{scope}[scale=\scale, shift={(\L*13.5,0)}]

\clip(4*\L-\cofset,2*\L-\cofset) rectangle (2*\M - 3*\L + \cofset, 2*\N - 3*\L + \cofset);

\tikzmath{
		{
			\filldraw[fill=lime] (5*\L,3*\L) rectangle ++(\L,2*\L);	
			\filldraw[fill=lime] (6*\L,4*\L) rectangle ++(\L,2*\L);
			\filldraw[fill=lime] (7*\L,3*\L) rectangle ++(\L,2*\L);	
			\filldraw[fill=lime] (8*\L,4*\L) rectangle ++(\L,2*\L);
			\filldraw[fill=lime] (9*\L,3*\L) rectangle ++(\L,2*\L);	
			\filldraw[fill=lime] (10*\L,4*\L) rectangle ++(\L,2*\L);	
			\filldraw[fill=lime] (11*\L,3*\L) rectangle ++(\L,2*\L);	
			\filldraw[fill=lime] (12*\L,4*\L) rectangle ++(\L,2*\L);		
			\filldraw[fill=lime] (13*\L,3*\L) rectangle ++(\L,2*\L);	
			\filldraw[fill=lime] (12*\L,4*\L) rectangle ++(\L,2*\L);		
		};
		{
			\filldraw[fill=lime] (13*\L,12*\L) rectangle ++(\L,2*\L);	
			\filldraw[fill=lime] (12*\L,11*\L) rectangle ++(\L,2*\L);	
			\filldraw[fill=lime] (11*\L,12*\L) rectangle ++(\L,2*\L);	
			\filldraw[fill=lime] (10*\L,11*\L) rectangle ++(\L,2*\L);	
			\filldraw[fill=lime] (9*\L,12*\L) rectangle ++(\L,2*\L);	
			\filldraw[fill=lime] (8*\L,11*\L) rectangle ++(\L,2*\L);	
			\filldraw[fill=lime] (7*\L,12*\L) rectangle ++(\L,2*\L);	
			\filldraw[fill=lime] (6*\L,11*\L) rectangle ++(\L,2*\L);	
			\filldraw[fill=lime] (5*\L,12*\L) rectangle ++(\L,2*\L);		
		};
		for \i in {1,...,2*\M}{
			{			
				\draw (\i*\L,-\L*0.5) -- (\i*\L,\L*2*\N-\L*0.5);
			};
		};
		for \i in {0,...,2*\N-1}{
			{			
				\draw (\L*0.5,\i*\L) -- (\L*2*\M+\L*0.5,\i*\L);
			};
		};
		for \i in {0,...,\M}{
			{
				\draw[color=gray, dashed] (2*\i*\L+0.5*\L,-\L*0.5) -- (2*\i*\L+0.5*\L,\L*2*\N-\L*0.5);
			};
		};
		for \i in {0,...,\N}{
			{			
				\draw[color=gray, dashed] (\L*0.5,2*\i*\L-0.5*\L) -- (\L*2*\M+\L*0.5,2*\i*\L-0.5*\L);
			};
		};
		for \x in {1,...,\M}{
			for \y in {0,...,\N-1}{
				{			
					\draw[fill=white] (2*\L*\x,2*\L*\y) circle[radius=0.05];
				};
			};
		};
		for \x in {0,...,\M-1}{
			for \y in {0,...,\N-1}{
				{			
					\draw[fill=white] (2*\L*\x+\L,2*\L*\y+\L) circle[radius=0.05];
				};
			};
		};
		for \x in {0,...,\M-1}{
			for \y in {0,...,\N-1}{
				{			
					\draw[fill] (2*\L*\x+\L,2*\L*\y) circle[radius=0.05];
				};
			};
		};	
		for \x in {1,...,\M}{
			for \y in {0,...,\N-1}{
				{			
					\draw[fill] (2*\L*\x,2*\L*\y+\L) circle[radius=0.05];
				};
			};
		};	
		for \x in {1,...,\M-1}{
			for \y in {1,...,\N-1}{
				{
					\tikzmath{\la=int(-\x+1);\lb=int(-\x+2);};		
					\node[scale=\fsize] at (2*\L*\x+0.5*\L,2*\L*\y-0.5*\L) {$_1$};
				};
			};
		};
		for \x in {0,...,\M-1}{
			for \y in {1,...,\N-1}{
				{
					\tikzmath{\la=int(\x-1);\lb=int(\x-1);};		
					\node[scale=\fsize] at (2*\L*\x+1.5*\L,2*\L*\y-0.5*\L) {$_2$};
				};
			};
		};
		for \x in {0,...,\M-1}{
			for \y in {0,...,\N-1}{
				{
					\tikzmath{\la=int(-\x+1);\lb=int(-\x+2);};
					\node[scale=\fsize] at (2*\L*\x+1.5*\L,2*\L*\y+\L-0.5*\L) {$_3$};
				};
			};
		};
		for \x in {1,...,\M-1}{
			for \y in {0,...,\N-1}{
				{
					\tikzmath{\la=int(\x-2);\lb=int(\x-2);};		
					\node[scale=\fsize] at (2*\L*\x+0.5*\L,2*\L*\y+\L-0.5*\L) {$_4$};
				};
			};
		};		
		for \y in {0,...,2}{
			for \x in {1,...,\M-2*\y-1}{
				{			
					\draw[color=blue, line width = \lwidth] (2*\L*\x+2*\L*\y,2*\L*\y)--(2*\L*\x+2*\L*\y+\L,2*\L*\y);
				};
			};
		};
		for \y in {0,...,1}{
			for \x in {1,...,\M-2*\y-2}{
				{			
					\draw[color=blue, line width = \lwidth] (2*\L*\x+2*\L*\y+\L,2*\L*\y+\L)--(2*\L*\x+2*\L*\y+2*\L,2*\L*\y+\L);
				};
			};
		};		
		for \y in {0,...,2}{
			for \x in {1,...,\M-2*\y-1}{
				{			
					\draw[color=purple, line width = \lwidth] (2*\L*\x+2*\L*\y,-2*\L*\y+2*\L*\N-\L)--(2*\L*\x+2*\L*\y+\L,-2*\L*\y+2*\L*\N-\L);
				};
			};
		};
		for \y in {0,...,1}{
			for \x in {1,...,\M-2*\y-2}{
				{			
					\draw[color=purple, line width = \lwidth] (2*\L*\x+2*\L*\y+\L,-2*\L*\y+2*\L*\N-2*\L)--(2*\L*\x+2*\L*\y+2*\L,-2*\L*\y+2*\L*\N-2*\L);
				};
			};
		};
		for \x in {1,...,3}{
			for \y in {0,...,\N-2*\x}{
				{			
					\draw[color=olive, line width = \lwidth] (2*\L*\x,2*\L*\y+2*\L*\x)--(2*\L*\x,2*\L*\y+2*\L*\x-\L);
				};
			};
		};
		for \x in {1,...,3}{
			for \y in {-1,...,\N-2*\x}{
				{			
					\draw[color=olive, line width = \lwidth] (2*\L*\x-\L,2*\L*\y+2*\L*\x+\L)--(2*\L*\x-\L,2*\L*\y+2*\L*\x);
				};
			};
		};
		for \x in {1,...,3}{
			for \y in {0,...,\N-2*\x}{
				{			
					\draw[color=orange, line width = \lwidth] (2*\L*\M-2*\L*\x+\L,2*\L*\y+2*\L*\x)--(2*\L*\M-2*\L*\x+\L,2*\L*\y+2*\L*\x-\L);
				};
			};
		};
		for \x in {1,...,3}{
			for \y in {-1,...,\N-2*\x}{
				{			
					\draw[color=orange, line width = \lwidth] (2*\L*\M-2*\L*\x+2*\L,2*\L*\y+2*\L*\x+\L)--(2*\L*\M-2*\L*\x+2*\L,2*\L*\y+2*\L*\x);
				};
			};
		};
		for \x in {2,...,\M-2}{
			for \y in {1,...,\N-5}{
				{			
					\draw[color=violet, line width = \lwidth] (2*\L+\x*\L+3*\L,4*\L+2*\y*\L)--(2*\L+\x*\L+3*\L,4*\L+2*\y*\L-\L);
				};
			};
		};		
};

\end{scope}
\begin{scope}[scale=\scale, shift={(0,-16*\L)}]

\clip(4*\L-\cofset,2*\L-\cofset) rectangle (2*\M - 3*\L + \cofset, 2*\N - 3*\L + \cofset);

\tikzmath{
		{
			\filldraw[fill=lime] (5*\L,3*\L) rectangle ++(\L,11*\L);	
			\filldraw[fill=lime] (6*\L,4*\L) rectangle ++(\L,9*\L);		
		};
		{
			\filldraw[fill=lime] (13*\L,3*\L) rectangle ++(\L,11*\L);	
			\filldraw[fill=lime] (12*\L,4*\L) rectangle ++(\L,9*\L);		
		};
		for \i in {1,...,2*\M}{
			{			
				\draw (\i*\L,-\L*0.5) -- (\i*\L,\L*2*\N-\L*0.5);
			};
		};
		for \i in {0,...,2*\N-1}{
			{			
				\draw (\L*0.5,\i*\L) -- (\L*2*\M+\L*0.5,\i*\L);
			};
		};
		for \i in {0,...,\M}{
			{
				\draw[color=gray, dashed] (2*\i*\L+0.5*\L,-\L*0.5) -- (2*\i*\L+0.5*\L,\L*2*\N-\L*0.5);
			};
		};
		for \i in {0,...,\N}{
			{			
				\draw[color=gray, dashed] (\L*0.5,2*\i*\L-0.5*\L) -- (\L*2*\M+\L*0.5,2*\i*\L-0.5*\L);
			};
		};
		for \x in {1,...,\M}{
			for \y in {0,...,\N-1}{
				{			
					\draw[fill=white] (2*\L*\x,2*\L*\y) circle[radius=0.05];
				};
			};
		};
		for \x in {0,...,\M-1}{
			for \y in {0,...,\N-1}{
				{			
					\draw[fill=white] (2*\L*\x+\L,2*\L*\y+\L) circle[radius=0.05];
				};
			};
		};
		for \x in {0,...,\M-1}{
			for \y in {0,...,\N-1}{
				{			
					\draw[fill] (2*\L*\x+\L,2*\L*\y) circle[radius=0.05];
				};
			};
		};	
		for \x in {1,...,\M}{
			for \y in {0,...,\N-1}{
				{			
					\draw[fill] (2*\L*\x,2*\L*\y+\L) circle[radius=0.05];
				};
			};
		};	
		for \x in {1,...,\M-1}{
			for \y in {1,...,\N-1}{
				{
					\tikzmath{\la=int(-\x+1);\lb=int(-\x+2);};		
					\node[scale=\fsize] at (2*\L*\x+0.5*\L,2*\L*\y-0.5*\L) {$_1$};
				};
			};
		};
		for \x in {0,...,\M-1}{
			for \y in {1,...,\N-1}{
				{
					\tikzmath{\la=int(\x-1);\lb=int(\x-1);};		
					\node[scale=\fsize] at (2*\L*\x+1.5*\L,2*\L*\y-0.5*\L) {$_2$};
				};
			};
		};
		for \x in {0,...,\M-1}{
			for \y in {0,...,\N-1}{
				{
					\tikzmath{\la=int(-\x+1);\lb=int(-\x+2);};
					\node[scale=\fsize] at (2*\L*\x+1.5*\L,2*\L*\y+\L-0.5*\L) {$_3$};
				};
			};
		};
		for \x in {1,...,\M-1}{
			for \y in {0,...,\N-1}{
				{
					\tikzmath{\la=int(\x-2);\lb=int(\x-2);};		
					\node[scale=\fsize] at (2*\L*\x+0.5*\L,2*\L*\y+\L-0.5*\L) {$_4$};
				};
			};
		};		
		for \y in {0,...,2}{
			for \x in {1,...,\M-2*\y-1}{
				{			
					\draw[color=blue, line width = \lwidth] (2*\L*\x+2*\L*\y,2*\L*\y)--(2*\L*\x+2*\L*\y+\L,2*\L*\y);
				};
			};
		};
		for \y in {0,...,1}{
			for \x in {1,...,\M-2*\y-2}{
				{			
					\draw[color=blue, line width = \lwidth] (2*\L*\x+2*\L*\y+\L,2*\L*\y+\L)--(2*\L*\x+2*\L*\y+2*\L,2*\L*\y+\L);
				};
			};
		};		
		for \y in {0,...,2}{
			for \x in {1,...,\M-2*\y-1}{
				{			
					\draw[color=purple, line width = \lwidth] (2*\L*\x+2*\L*\y,-2*\L*\y+2*\L*\N-\L)--(2*\L*\x+2*\L*\y+\L,-2*\L*\y+2*\L*\N-\L);
				};
			};
		};
		for \y in {0,...,1}{
			for \x in {1,...,\M-2*\y-2}{
				{			
					\draw[color=purple, line width = \lwidth] (2*\L*\x+2*\L*\y+\L,-2*\L*\y+2*\L*\N-2*\L)--(2*\L*\x+2*\L*\y+2*\L,-2*\L*\y+2*\L*\N-2*\L);
				};
			};
		};
		for \x in {1,...,3}{
			for \y in {0,...,\N-2*\x}{
				{			
					\draw[color=olive, line width = \lwidth] (2*\L*\x,2*\L*\y+2*\L*\x)--(2*\L*\x,2*\L*\y+2*\L*\x-\L);
				};
			};
		};
		for \x in {1,...,3}{
			for \y in {-1,...,\N-2*\x}{
				{			
					\draw[color=olive, line width = \lwidth] (2*\L*\x-\L,2*\L*\y+2*\L*\x+\L)--(2*\L*\x-\L,2*\L*\y+2*\L*\x);
				};
			};
		};
		for \x in {1,...,3}{
			for \y in {0,...,\N-2*\x}{
				{			
					\draw[color=orange, line width = \lwidth] (2*\L*\M-2*\L*\x+\L,2*\L*\y+2*\L*\x)--(2*\L*\M-2*\L*\x+\L,2*\L*\y+2*\L*\x-\L);
				};
			};
		};
		for \x in {1,...,3}{
			for \y in {-1,...,\N-2*\x}{
				{			
					\draw[color=orange, line width = \lwidth] (2*\L*\M-2*\L*\x+2*\L,2*\L*\y+2*\L*\x+\L)--(2*\L*\M-2*\L*\x+2*\L,2*\L*\y+2*\L*\x);
				};
			};
		};
		for \x in {2,...,\M-2}{
			for \y in {1,...,\N-5}{
				{			
					\draw[color=violet, line width = \lwidth] (2*\L+\x*\L+3*\L,4*\L+2*\y*\L)--(2*\L+\x*\L+3*\L,4*\L+2*\y*\L-\L);
				};
			};
		};		
};

\end{scope}
\begin{scope}[scale=\scale, shift={(\L*13.5,-16*\L)}]

\clip(4*\L-\cofset,2*\L-\cofset) rectangle (2*\M - 3*\L + \cofset, 2*\N - 3*\L + \cofset);

\tikzmath{
		{
			\filldraw[fill=lime] (5*\L,5*\L) rectangle ++(9*\L,7*\L);
		};
		{
			\filldraw[fill=lime] (5*\L,3*\L) rectangle ++(\L,2*\L);	
			\filldraw[fill=lime] (6*\L,4*\L) rectangle ++(\L,2*\L);
			\filldraw[fill=lime] (7*\L,3*\L) rectangle ++(\L,2*\L);	
			\filldraw[fill=lime] (8*\L,4*\L) rectangle ++(\L,2*\L);
			\filldraw[fill=lime] (9*\L,3*\L) rectangle ++(\L,2*\L);	
			\filldraw[fill=lime] (10*\L,4*\L) rectangle ++(\L,2*\L);	
			\filldraw[fill=lime] (11*\L,3*\L) rectangle ++(\L,2*\L);	
			\filldraw[fill=lime] (12*\L,4*\L) rectangle ++(\L,2*\L);		
			\filldraw[fill=lime] (13*\L,3*\L) rectangle ++(\L,2*\L);	
			\filldraw[fill=lime] (12*\L,4*\L) rectangle ++(\L,2*\L);		
		};
		{
			\filldraw[fill=lime] (13*\L,12*\L) rectangle ++(\L,2*\L);	
			\filldraw[fill=lime] (12*\L,11*\L) rectangle ++(\L,2*\L);	
			\filldraw[fill=lime] (11*\L,12*\L) rectangle ++(\L,2*\L);	
			\filldraw[fill=lime] (10*\L,11*\L) rectangle ++(\L,2*\L);	
			\filldraw[fill=lime] (9*\L,12*\L) rectangle ++(\L,2*\L);	
			\filldraw[fill=lime] (8*\L,11*\L) rectangle ++(\L,2*\L);	
			\filldraw[fill=lime] (7*\L,12*\L) rectangle ++(\L,2*\L);	
			\filldraw[fill=lime] (6*\L,11*\L) rectangle ++(\L,2*\L);	
			\filldraw[fill=lime] (5*\L,12*\L) rectangle ++(\L,2*\L);		
		};
		for \i in {1,...,2*\M}{
			{			
				\draw (\i*\L,-\L*0.5) -- (\i*\L,\L*2*\N-\L*0.5);
			};
		};
		for \i in {0,...,2*\N-1}{
			{			
				\draw (\L*0.5,\i*\L) -- (\L*2*\M+\L*0.5,\i*\L);
			};
		};
		for \i in {0,...,\M}{
			{
				\draw[color=gray, dashed] (2*\i*\L+0.5*\L,-\L*0.5) -- (2*\i*\L+0.5*\L,\L*2*\N-\L*0.5);
			};
		};
		for \i in {0,...,\N}{
			{			
				\draw[color=gray, dashed] (\L*0.5,2*\i*\L-0.5*\L) -- (\L*2*\M+\L*0.5,2*\i*\L-0.5*\L);
			};
		};
		for \x in {1,...,\M}{
			for \y in {0,...,\N-1}{
				{			
					\draw[fill=white] (2*\L*\x,2*\L*\y) circle[radius=0.05];
				};
			};
		};
		for \x in {0,...,\M-1}{
			for \y in {0,...,\N-1}{
				{			
					\draw[fill=white] (2*\L*\x+\L,2*\L*\y+\L) circle[radius=0.05];
				};
			};
		};
		for \x in {0,...,\M-1}{
			for \y in {0,...,\N-1}{
				{			
					\draw[fill] (2*\L*\x+\L,2*\L*\y) circle[radius=0.05];
				};
			};
		};	
		for \x in {1,...,\M}{
			for \y in {0,...,\N-1}{
				{			
					\draw[fill] (2*\L*\x,2*\L*\y+\L) circle[radius=0.05];
				};
			};
		};	
		for \x in {1,...,\M-1}{
			for \y in {1,...,\N-1}{
				{
					\tikzmath{\la=int(-\x+1);\lb=int(-\x+2);};		
					\node[scale=\fsize] at (2*\L*\x+0.5*\L,2*\L*\y-0.5*\L) {$_1$};
				};
			};
		};
		for \x in {0,...,\M-1}{
			for \y in {1,...,\N-1}{
				{
					\tikzmath{\la=int(\x-1);\lb=int(\x-1);};		
					\node[scale=\fsize] at (2*\L*\x+1.5*\L,2*\L*\y-0.5*\L) {$_2$};
				};
			};
		};
		for \x in {0,...,\M-1}{
			for \y in {0,...,\N-1}{
				{
					\tikzmath{\la=int(-\x+1);\lb=int(-\x+2);};
					\node[scale=\fsize] at (2*\L*\x+1.5*\L,2*\L*\y+\L-0.5*\L) {$_3$};
				};
			};
		};
		for \x in {1,...,\M-1}{
			for \y in {0,...,\N-1}{
				{
					\tikzmath{\la=int(\x-2);\lb=int(\x-2);};		
					\node[scale=\fsize] at (2*\L*\x+0.5*\L,2*\L*\y+\L-0.5*\L) {$_4$};
				};
			};
		};		
		for \y in {0,...,2}{
			for \x in {1,...,\M-2*\y-1}{
				{			
					\draw[color=blue, line width = \lwidth] (2*\L*\x+2*\L*\y,2*\L*\y)--(2*\L*\x+2*\L*\y+\L,2*\L*\y);
				};
			};
		};
		for \y in {0,...,1}{
			for \x in {1,...,\M-2*\y-2}{
				{			
					\draw[color=blue, line width = \lwidth] (2*\L*\x+2*\L*\y+\L,2*\L*\y+\L)--(2*\L*\x+2*\L*\y+2*\L,2*\L*\y+\L);
				};
			};
		};		
		for \y in {0,...,2}{
			for \x in {1,...,\M-2*\y-1}{
				{			
					\draw[color=purple, line width = \lwidth] (2*\L*\x+2*\L*\y,-2*\L*\y+2*\L*\N-\L)--(2*\L*\x+2*\L*\y+\L,-2*\L*\y+2*\L*\N-\L);
				};
			};
		};
		for \y in {0,...,1}{
			for \x in {1,...,\M-2*\y-2}{
				{			
					\draw[color=purple, line width = \lwidth] (2*\L*\x+2*\L*\y+\L,-2*\L*\y+2*\L*\N-2*\L)--(2*\L*\x+2*\L*\y+2*\L,-2*\L*\y+2*\L*\N-2*\L);
				};
			};
		};
		for \x in {1,...,3}{
			for \y in {0,...,\N-2*\x}{
				{			
					\draw[color=olive, line width = \lwidth] (2*\L*\x,2*\L*\y+2*\L*\x)--(2*\L*\x,2*\L*\y+2*\L*\x-\L);
				};
			};
		};
		for \x in {1,...,3}{
			for \y in {-1,...,\N-2*\x}{
				{			
					\draw[color=olive, line width = \lwidth] (2*\L*\x-\L,2*\L*\y+2*\L*\x+\L)--(2*\L*\x-\L,2*\L*\y+2*\L*\x);
				};
			};
		};
		for \x in {1,...,3}{
			for \y in {0,...,\N-2*\x}{
				{			
					\draw[color=orange, line width = \lwidth] (2*\L*\M-2*\L*\x+\L,2*\L*\y+2*\L*\x)--(2*\L*\M-2*\L*\x+\L,2*\L*\y+2*\L*\x-\L);
				};
			};
		};
		for \x in {1,...,3}{
			for \y in {-1,...,\N-2*\x}{
				{			
					\draw[color=orange, line width = \lwidth] (2*\L*\M-2*\L*\x+2*\L,2*\L*\y+2*\L*\x+\L)--(2*\L*\M-2*\L*\x+2*\L,2*\L*\y+2*\L*\x);
				};
			};
		};
		for \x in {2,...,\M-2}{
			for \y in {1,...,\N-5}{
				{			
					\draw[color=violet, line width = \lwidth] (2*\L+\x*\L+3*\L,4*\L+2*\y*\L)--(2*\L+\x*\L+3*\L,4*\L+2*\y*\L-\L);
				};
			};
		};		
};

\end{scope}

\end{tikzpicture}
}
\caption{Toda bipartite lattices with ``empty room'' configuration $D_0$ drawn. Faces involved in the rotations corresponding to addition of boxes weighted by $Q_0, Q_{1,B},Q_{1,F}, Q_{2}$ are highlighted by lime colour.}
\label{fig:gridTodaWithDimers}
\end{figure}
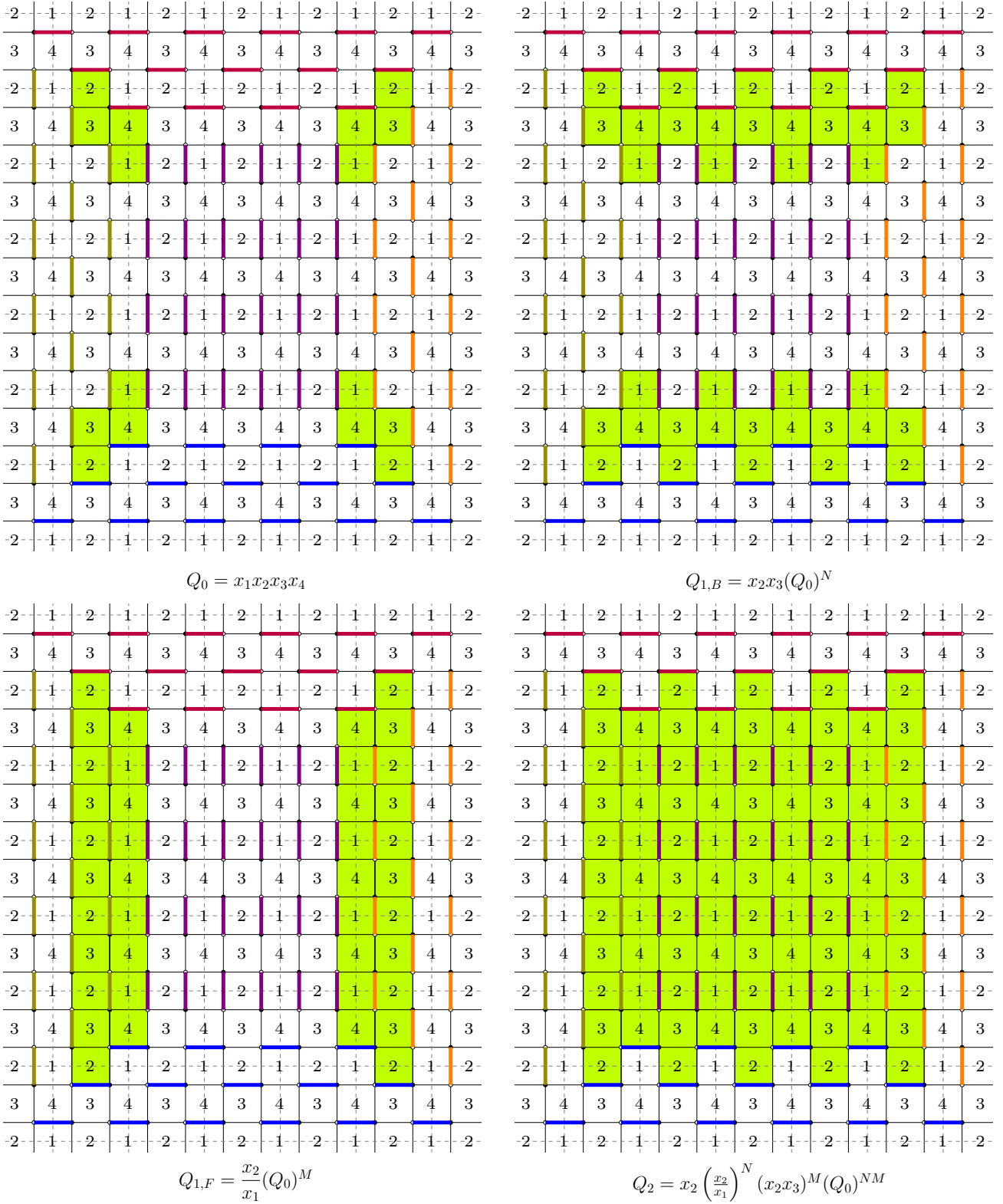

The starting point for the box counting combinatorics is the ``empty room'' dimers configuration, which is drawn on all four panels of Fig.~\ref{fig:gridTodaWithDimers} by coloured dimers. The structure of configuration is similar to the structure of amoeba drawn on Fig.~\ref{fig:amoebaToda}: there are four unbounded domains corresponding to $\Dom_1,\Dom_2,\Dom_3, \Dom_4$, and one internal domain $\Dom_0$. Dimers configurations in unbounded domains are just the tilings by configurations corresponding to four ``external'' monomials at $\lambda,\lambda^{-1},\mu,\mu^{-1}$ in (\ref{eq:curveToda}), and configuration in $\Dom_0$ is one of those at $\lambda^{0} \mu^{0}$. Two parameters defining this configuration are width $N$ and height $M$ of central domain. For the configuration on Fig. \ref{fig:gridTodaWithDimers} we have $N=4$, $M=5$ by the number of fundamental domains filled by purple dimers plus 1.

\begin{figure}[!h]
\centering
\scalebox{0.65}{
\begin{tikzpicture}

\tikzmath{
	\N=9;\M=9;
	\scale=1;
	\cofset=0.8;
	\L=1;
	\fsize=1.75;
	\lwidth=1mm;
}

\begin{scope}[scale=\scale]

\clip(4*\L-\cofset,2*\L-\cofset) rectangle (2*\M - 3*\L + \cofset, 2*\N - 3*\L + \cofset);
\tikzmath{
		{
			\filldraw [fill=lime] (4*\L,1*\L) -- ++(0,\L) -- ++(\L,0) -- ++(0,\L) -- ++(\L,0) -- ++(0,\L) -- ++(\L,0) -- ++(0,-\L) -- ++(\L,0) -- ++(0,\L) -- ++(\L,0) -- ++(0,-\L) -- ++(\L,0) -- ++(0,\L) -- ++(\L,0) -- ++(0,-\L) -- ++(\L,0) -- ++(0,\L) -- ++(\L,0) -- ++(0,-\L) -- ++(\L,0) -- ++(0,-\L) -- ++(\L,0) -- ++(0,-\L);
		};
		{
			\filldraw[fill=lime] (16*\L,14*\L) -- ++(-\L,0) -- ++(0,-\L) -- ++(-\L,0) -- ++(0,-\L) -- ++(-\L,0) -- ++(0,-\L) -- ++(\L,0) -- ++(0,-\L) -- ++(-\L,0) -- ++(0,-\L) -- ++(\L,0) -- ++(0,-\L) -- ++(-\L,0) -- ++(0,-\L) -- ++(\L,0) -- ++(0,-\L) -- ++(-\L,0) -- ++(0,-\L) -- ++(\L,0) -- ++(0,-\L) -- ++(\L,0) -- ++(0,-\L) -- ++(\L,0);
		};
		for \i in {1,...,2*\M}{
			{			
				\draw (\i*\L,-\L*0.5) -- (\i*\L,\L*2*\N-\L*0.5);
			};
		};
		for \i in {0,...,2*\N-1}{
			{			
				\draw (\L*0.5,\i*\L) -- (\L*2*\M+\L*0.5,\i*\L);
			};
		};
		for \i in {0,...,\M}{
			{
				\draw[color=gray, dashed] (2*\i*\L+0.5*\L,-\L*0.5) -- (2*\i*\L+0.5*\L,\L*2*\N-\L*0.5);
			};
		};
		for \i in {0,...,\N}{
			{			
				\draw[color=gray, dashed] (\L*0.5,2*\i*\L-0.5*\L) -- (\L*2*\M+\L*0.5,2*\i*\L-0.5*\L);
			};
		};
		for \x in {1,...,\M}{
			for \y in {0,...,\N-1}{
				{			
					\draw[fill=white] (2*\L*\x,2*\L*\y) circle[radius=0.05];
				};
			};
		};
		for \x in {0,...,\M-1}{
			for \y in {0,...,\N-1}{
				{			
					\draw[fill=white] (2*\L*\x+\L,2*\L*\y+\L) circle[radius=0.05];
				};
			};
		};
		for \x in {0,...,\M-1}{
			for \y in {0,...,\N-1}{
				{			
					\draw[fill] (2*\L*\x+\L,2*\L*\y) circle[radius=0.05];
				};
			};
		};	
		for \x in {1,...,\M}{
			for \y in {0,...,\N-1}{
				{			
					\draw[fill] (2*\L*\x,2*\L*\y+\L) circle[radius=0.05];
				};
			};
		};	
		for \x in {1,...,\M-1}{
			for \y in {1,...,\N-1}{
				{
					\tikzmath{\la=int(-\x+1);\lb=int(-\x+2);};		
					\node[scale=\fsize] at (2*\L*\x+0.5*\L,2*\L*\y-0.5*\L) {$_1$};
				};
			};
		};
		for \x in {0,...,\M-1}{
			for \y in {1,...,\N-1}{
				{
					\tikzmath{\la=int(\x-1);\lb=int(\x-1);};		
					\node[scale=\fsize] at (2*\L*\x+1.5*\L,2*\L*\y-0.5*\L) {$_2$};
				};
			};
		};
		for \x in {0,...,\M-1}{
			for \y in {0,...,\N-1}{
				{
					\tikzmath{\la=int(-\x+1);\lb=int(-\x+2);};
					\node[scale=\fsize] at (2*\L*\x+1.5*\L,2*\L*\y+\L-0.5*\L) {$_3$};
				};
			};
		};
		for \x in {1,...,\M-1}{
			for \y in {0,...,\N-1}{
				{
					\tikzmath{\la=int(\x-2);\lb=int(\x-2);};		
					\node[scale=\fsize] at (2*\L*\x+0.5*\L,2*\L*\y+\L-0.5*\L) {$_4$};
				};
			};
		};		
		for \y in {0,...,2}{
			for \x in {1,...,\M-2*\y-1}{
				{			
					\draw[color=blue, line width = \lwidth] (2*\L*\x+2*\L*\y,2*\L*\y)--(2*\L*\x+2*\L*\y+\L,2*\L*\y);
				};
			};
		};
		for \y in {0,...,1}{
			for \x in {1,...,\M-2*\y-2}{
				{			
					\draw[color=blue, line width = \lwidth] (2*\L*\x+2*\L*\y+\L,2*\L*\y+\L)--(2*\L*\x+2*\L*\y+2*\L,2*\L*\y+\L);
				};
			};
		};		
		for \y in {0,...,2}{
			for \x in {1,...,\M-2*\y-1}{
				{			
					\draw[color=purple, line width = \lwidth] (2*\L*\x+2*\L*\y,-2*\L*\y+2*\L*\N-\L)--(2*\L*\x+2*\L*\y+\L,-2*\L*\y+2*\L*\N-\L);
				};
			};
		};
		for \y in {0,...,1}{
			for \x in {1,...,\M-2*\y-2}{
				{			
					\draw[color=purple, line width = \lwidth] (2*\L*\x+2*\L*\y+\L,-2*\L*\y+2*\L*\N-2*\L)--(2*\L*\x+2*\L*\y+2*\L,-2*\L*\y+2*\L*\N-2*\L);
				};
			};
		};
		for \x in {1,...,3}{
			for \y in {0,...,\N-2*\x}{
				{			
					\draw[color=olive, line width = \lwidth] (2*\L*\x,2*\L*\y+2*\L*\x)--(2*\L*\x,2*\L*\y+2*\L*\x-\L);
				};
			};
		};
		for \x in {1,...,3}{
			for \y in {-1,...,\N-2*\x}{
				{			
					\draw[color=olive, line width = \lwidth] (2*\L*\x-\L,2*\L*\y+2*\L*\x+\L)--(2*\L*\x-\L,2*\L*\y+2*\L*\x);
				};
			};
		};
		for \x in {1,...,3}{
			for \y in {0,...,\N-2*\x}{
				{			
					\draw[color=orange, line width = \lwidth] (2*\L*\M-2*\L*\x+\L,2*\L*\y+2*\L*\x)--(2*\L*\M-2*\L*\x+\L,2*\L*\y+2*\L*\x-\L);
				};
			};
		};
		for \x in {1,...,3}{
			for \y in {-1,...,\N-2*\x}{
				{			
					\draw[color=orange, line width = \lwidth] (2*\L*\M-2*\L*\x+2*\L,2*\L*\y+2*\L*\x+\L)--(2*\L*\M-2*\L*\x+2*\L,2*\L*\y+2*\L*\x);
				};
			};
		};
		for \x in {2,...,\M-2}{
			for \y in {1,...,\N-5}{
				{			
					\draw[color=violet, line width = \lwidth] (2*\L+\x*\L+3*\L,4*\L+2*\y*\L)--(2*\L+\x*\L+3*\L,4*\L+2*\y*\L-\L);
				};
			};
		};		
};

\end{scope}
\begin{scope}[scale=\scale, shift={(\L*13.5,0)}]

\clip(4*\L-\cofset,2*\L-\cofset) rectangle (2*\M - 3*\L + \cofset, 2*\N - 3*\L + \cofset);

\tikzmath{
		{
			\filldraw[fill=lime] (5*\L,3*\L) rectangle ++(\L,3*\L);		
		};
		{
			\filldraw[fill=lime] (8*\L,7*\L) rectangle ++(\L,3*\L);		
		};
		{
			\filldraw[fill=lime] (7*\L,3*\L) rectangle ++(\L,\L);		
		};
		{
			\filldraw[fill=lime] (6*\L,7*\L) rectangle ++(\L,\L);		
		};
		{
			\filldraw[fill=lime] (10*\L,3*\L) rectangle ++(\L,3*\L);
			\filldraw[fill=lime] (11*\L,4*\L) rectangle ++(\L,\L);
			\filldraw[fill=lime] (12*\L,3*\L) rectangle ++(\L,3*\L);		
		};
		{
			\filldraw[fill=lime] (11*\L,7*\L) rectangle ++(\L,3*\L);
			\filldraw[fill=lime] (12*\L,8*\L) rectangle ++(\L,\L);
			\filldraw[fill=lime] (13*\L,7*\L) rectangle ++(\L,5*\L);		
		};
		{
			\filldraw[fill=lime] (6*\L,11*\L) rectangle ++(\L,3*\L);
			\filldraw[fill=lime] (7*\L,12*\L) rectangle ++(\L,\L);
			\filldraw[fill=lime] (8*\L,11*\L) rectangle ++(\L,3*\L);
			\filldraw[fill=lime] (9*\L,12*\L) rectangle ++(\L,\L);
			\filldraw[fill=lime] (10*\L,11*\L) rectangle ++(\L,3*\L);		
		};
		for \i in {1,...,2*\M}{
			{			
				\draw (\i*\L,-\L*0.5) -- (\i*\L,\L*2*\N-\L*0.5);
			};
		};
		for \i in {0,...,2*\N-1}{
			{			
				\draw (\L*0.5,\i*\L) -- (\L*2*\M+\L*0.5,\i*\L);
			};
		};
		for \i in {0,...,\M}{
			{
				\draw[color=gray, dashed] (2*\i*\L+0.5*\L,-\L*0.5) -- (2*\i*\L+0.5*\L,\L*2*\N-\L*0.5);
			};
		};
		for \i in {0,...,\N}{
			{			
				\draw[color=gray, dashed] (\L*0.5,2*\i*\L-0.5*\L) -- (\L*2*\M+\L*0.5,2*\i*\L-0.5*\L);
			};
		};
		for \x in {1,...,\M}{
			for \y in {0,...,\N-1}{
				{			
					\draw[fill=white] (2*\L*\x,2*\L*\y) circle[radius=0.05];
				};
			};
		};
		for \x in {0,...,\M-1}{
			for \y in {0,...,\N-1}{
				{			
					\draw[fill=white] (2*\L*\x+\L,2*\L*\y+\L) circle[radius=0.05];
				};
			};
		};
		for \x in {0,...,\M-1}{
			for \y in {0,...,\N-1}{
				{			
					\draw[fill] (2*\L*\x+\L,2*\L*\y) circle[radius=0.05];
				};
			};
		};	
		for \x in {1,...,\M}{
			for \y in {0,...,\N-1}{
				{			
					\draw[fill] (2*\L*\x,2*\L*\y+\L) circle[radius=0.05];
				};
			};
		};	
		for \x in {1,...,\M-1}{
			for \y in {1,...,\N-1}{
				{
					\tikzmath{\la=int(-\x+1);\lb=int(-\x+2);};		
					\node[scale=\fsize] at (2*\L*\x+0.5*\L,2*\L*\y-0.5*\L) {$_1$};
				};
			};
		};
		for \x in {0,...,\M-1}{
			for \y in {1,...,\N-1}{
				{
					\tikzmath{\la=int(\x-1);\lb=int(\x-1);};		
					\node[scale=\fsize] at (2*\L*\x+1.5*\L,2*\L*\y-0.5*\L) {$_2$};
				};
			};
		};
		for \x in {0,...,\M-1}{
			for \y in {0,...,\N-1}{
				{
					\tikzmath{\la=int(-\x+1);\lb=int(-\x+2);};
					\node[scale=\fsize] at (2*\L*\x+1.5*\L,2*\L*\y+\L-0.5*\L) {$_3$};
				};
			};
		};
		for \x in {1,...,\M-1}{
			for \y in {0,...,\N-1}{
				{
					\tikzmath{\la=int(\x-2);\lb=int(\x-2);};		
					\node[scale=\fsize] at (2*\L*\x+0.5*\L,2*\L*\y+\L-0.5*\L) {$_4$};
				};
			};
		};		
		for \x in {0,...,2*\M}{
			for \y in {0,...,\N}{
				{			
					\draw[color=violet, line width = \lwidth] (\x*\L,2*\y*\L)--(\x*\L,2*\y*\L-\L);
				};
			};
		};		
};

\end{scope}

\end{tikzpicture}
}
\caption{Left: ``Unbounded'' rotations, changing $(N,M) \mapsto (N+1, M+1)$ and $(N,M)\mapsto (N,M+1)$. Right: different types of rotation possible in the central region, which are freezing out in tropical limit. The weights of rotations shown on picture by lime color are $x_1, (x_1)^2 x_4, (x_2)^{-1}, (x_2)^{-2} (x_3)^{-1}, (x_1)^4 x_3 (x_4)^2, (x_1)^6 (x_3)^2 (x_4)^3, (x_2)^{-5} (x_3)^{-3} (x_4)^{-1},...$}
\label{fig:gridTodaWithDimers2}
\end{figure}
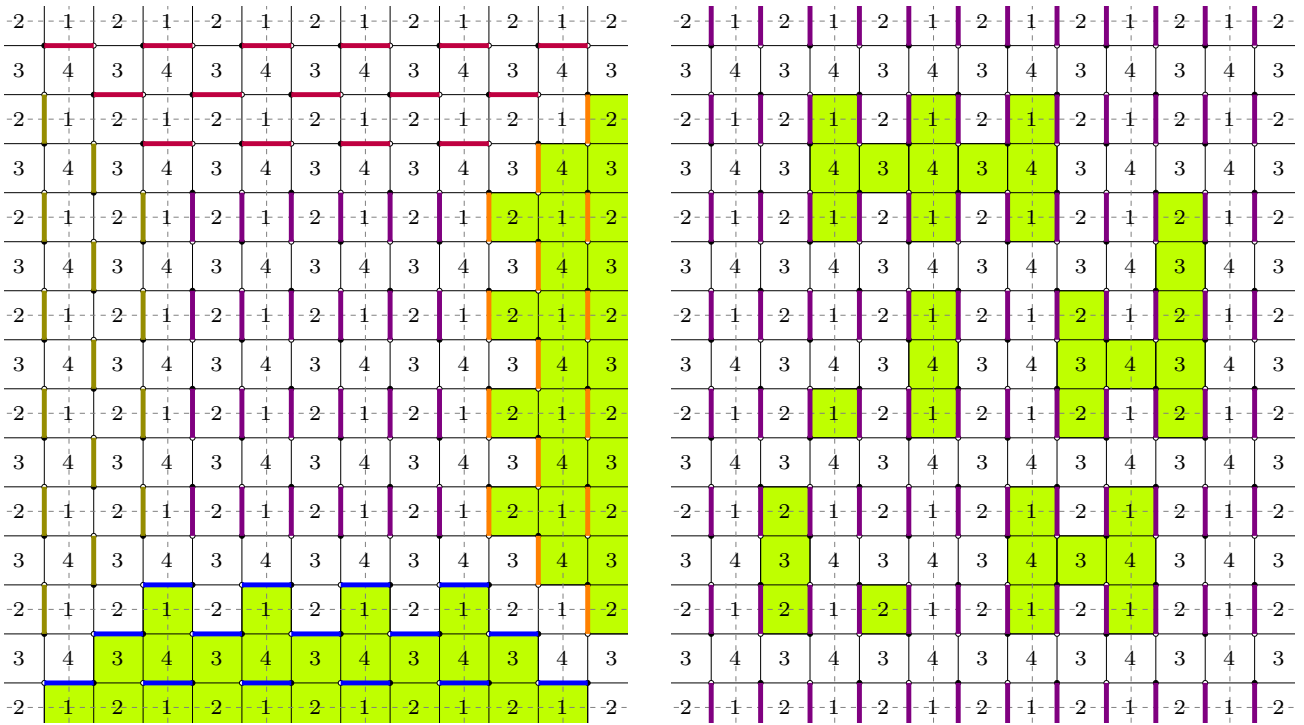

The ``rotation in the set of faces'' is a transition from one dimers configuration to another by choosing such a set of faces that exactly half of edges on their common boundary (each second edge) is contained in dimers configuration, and exchanging sets of occupied and non-occupied edges on this boundary. This changes the weight of the dimers configuration by the product of the corresponding face weights. There are four classes of transformations of the ``empty room'' configuration (and configurations obtained from it by these transformations), which correspond to adding of different types of boxes to the room:
\begin{itemize}
\item Four rotation in the sets of faces as on Fig.~\ref{fig:gridTodaWithDimers}, left, top. Each rotation of this type is weighted by $q=Q_0 = x_1 x_2 x_3 x_4$, and corresponds to the addition of $0d$ box to one of four $3d$ Young diagrams located in the corners of the room. First rotation of this type opens possibility for three more similar rotations in the adjacent locations, which is in agreement with the fact that there are three $3d$ Young diagrams containing two boxes. Similar matching works further, until the diagram growing in one corner touches diagram from another corner. This can be easily seen considering e.g. left top corner of the ``room'' and erasing edges between the faces $2$ and $3$, $3$ and $4$, $4$ and $1$, which are not covered by any dimers there and are not involved into transformations then. Making reduction of pairs of adjacent $2$-valent vertices of bipartite graph after erasing, we get hexagonal lattices, which provides $3d$ box counting \cite{ORV03}.
\item Two rotations weighted by $Q_{1,B}$ as shown on a top right panel, and two ones weighted by $Q_{1,F}$ from a bottom left panel are corresponding to addition of $1d$ boxes constituting four $2d$ Young diagrams. These $2d$ Young diagrams can be considered as a so long lines of boxes added to the corners, that they meet each other. However, since the shapes corresponding to addition of boxes to different corners are different, there is a mismatch, because of which $Q_{1,B}$ and $Q_{1,F}$ are not simply degrees of $q$, but contain also other combinations of the weight of faces. So the $2d$ Young diagrams determine the initial shape, on the top of which $3d$ Young diagrams are built.
\item Rotation shown on a bottom right panel is weighted by $Q_2$ and results in the change $(N,M) \mapsto (N+2,M+2)$. In terms of the boxes, this can be viewed as change of the level of ``floor'' in the room. Since you can repeatedly apply this transformations, they are enumerated by $\mathbb{N}$ or $1d$ Young diagrams. 
\item There are also two types of transformations of infinite weights, shown on Fig.~\ref{fig:gridTodaWithDimers2}, left. They change $(N,M) \mapsto (N+1,M)$ and $(N,M) \mapsto (N,M+1)$, and do  not contribute to the partition, since we assume boundary conditions at infinity to be fixed. However, we will be back to them in the Discussion section, we expect them to play an important role in the context of solutions of $q$-difference equations with the partition functions of dimers. From the point of view of box counting, these transformations are corresponding to shifts of the ``walls'' of the room. 
\end{itemize}
Summation of $3d$ and $2d$ boxes is given by $Z_{\mathrm{boxes}}(q,Q_B, Q_F)$ in (\ref{eq:ZinstD}), $Q_2, Q_B$ and $Q_F$ in the formula are taken at some large fixed values of $(N,M)$. The weight in front of $Z_{\mathrm{boxes}}(q,Q_B, Q_F)$ originates from multiplication by $Q_2$ factors for $(N,M)$, $(N+2,M+2)$,...,$(N+2n-2,M+2n-2)$.  The growth rate $\tfrac{4}{3}\db n^3$ in the exponent is related to the volume of pyramid. It matches nicely with the leading in $u$ term
\begin{equation}
\prep \sim -\dfrac{4}{3} \dfrac{(\R5 u)^3}{2\pi \ii}
\end{equation}
in (\ref{eq:prepTrop}), where $2\pi \ii$ comes from the different normalization of prepotential compared to the volume. The external summation over $n$ is for the summation over the ``heights'' of the floor, or divergences of size of central domain from $(N,M)$. It has to go in the limits $-\mathrm{min}(N,M) \leq n \leq +\infty$, but we can take it to be two-sided infinite, since we are working in approximation $N,M \to +\infty$, which is also important for $3d$ Young diagrams to not to touch each other.

\subsection{Inconsistency of ``freezing out'' and thermodynamic limit}
\label{ss:problems}
We are going to suggest now how to freeze all non-boxcounting ``rotations'' at once by the proper tuning of weights of faces, and show then why thermodynamically this is incompatible with $N,M\to +\infty$ limit.

First of all, there are no possible local rotations of size $\ll N,M$ in non-bounded domains $\Dom_1, \Dom_2, \Dom_3, \Dom_4$, since the dimers configurations which tile them are ``extremal'': the difference with any other configuration will be a collection of paths which go in one direction and can't go back. There are many possible local rotations in the central domain, as it is shown on Fig.~\ref{fig:gridTodaWithDimers2}, right. We are looking for such limit of faces' weights to zeroes or infinities (tropical limit), that weights of all rotations in this domain are suppressed. We also want to keep finite $q$, so we will assume now $x_1 x_2 x_3 x_4=1$ in compare with the weights of individual faces. Then, the partition function of local rotations can be estimated, by selecting the term at $\lambda^0 \mu^0$ in the partition function on large torus of size $L \times L$ \cite{KOS03}, which can be estimated as
\begin{equation}
\Z_{\mathbb{T}^2, L \times L}(\Gamma, w; D_0)|_{\lambda^0 \mu^0} \leq (w_3 w_5)^{-L^2} \prod_{a=1}^{L} \prod_{b=1}^{L} \det \K_1(\lambda e^{\frac{2\pi \ii a}{L}}, \mu e^{\frac{2\pi \ii b}{L}})|_{\lambda^0 \mu^0} \leq \left( \dfrac{\det \K_1(\lambda, \mu)}{w_3 w_5} \right)^{L^2}|_{\lambda^0 \mu^0} = 
\end{equation}
$$
=\sum\limits_{2a+2b+c=L^2} \left(\dfrac{x_1}{x_2}\right)^a (x_1 x_4)^{b} \left(1 + x_1 + x_1 x_4 + x_1 x_3 x_4\right)^c
$$
Using additive variables $\xi_i$ in $x_i = e^{\R5 \xi_i + \we{x}_i}$ at $\R5 \to + \infty$, all terms except $1$ are vanishing if
\begin{equation}
\label{eq:constrTrop}
\xi_1<0, ~~~ \xi_1 + \xi_4 < 0, ~~~ \xi_1 +\xi_3 + \xi_4 < 0, ~~~ \xi_1 + \xi_2 + \xi_3 + \xi_4 = 0.
\end{equation} 
As a check, one can see that all of the rotations shown on Fig.~\ref{fig:gridTodaWithDimers2}, right, are suppressed in this limit. It also has to be shown that these bounds are enough to suppress all the local rotations in between of domain $\Dom_0$ and other domains $\Dom_i$. We do not know how to show this systematically though.\\

Unfortunately, constraints (\ref{eq:constrTrop}) are inconsistent with the thermodynamic limit $N,M\to +\infty$. We require that in thermodynamic limit all the weights $Q_0, Q_{1,B}, Q_{1,F}, Q_2$ should be finite, not becoming $0$ or $\infty$. Inverting formulas for their weights on Fig.~\ref{fig:gridTodaWithDimers}, one gets
\begin{equation}
\label{eq:todaClustersByQ}
x_1 = X_{N+1,M},~~~
x_2 = X_{N,M},~~~
x_3 = \dfrac{1}{X_{N,M+1}},~~~
x_4 = \dfrac{1}{X_{N+1,M-1}},~~~\text{where}~~~
X_{N,M} = \dfrac{Q_2 \cdot (Q_0)^{NM}}{(Q_{1,B})^{M} \cdot  (Q_{1,F})^{N}}.
\end{equation}
The leading terms are determined here by $Q_0 = q = e^{-\db}$ since $NM \gg N,M \gg 0$, so taking $\R5=NM$, one gets
\begin{equation}
\xi_1 = -\db, ~~~ \xi_2 = -\db, ~~~ \xi_3 = \db, ~~~ \xi_4 = \db ~~~ \Rightarrow ~~~ \xi_1 + \xi_3 + \xi_4 = \db > 0,
\end{equation}
which is inconsistent with (\ref{eq:constrTrop}).

Another issue with thermodynamic limit is the instability due to the multiplier $\sim q^{\tfrac{4}{3}n^3}$ in (\ref{eq:ZinstD}). Even if all $Q$ are finite and non-boxcounting degrees of freedom are suppressed, the cubic term at $n\to -\infty$ dominates all the other contributions at fixed $n$, making small $n$ preferable and breaking $N,M \gg n \gg 1$ assumptions.

\section{Discussion}
\label{s:discussion}

In the paper we made several steps towards understanding the role of cluster algebras in the theory of topological string. We have shown how starting from the ``deautonomization'' of cluster integrable system one naturally gets objects related to topological string: either Seiberg-Witten prepotential in the ``melting'' limit, or boxcounting of topological vertices in the ``tropical'' limit. Despite of inconsistencies, outlined in the Section \ref{ss:problems}, this consideration seems to provide proper framework for the construction of the arrow shown on Fig.~\ref{fig:mainDiagram} in the Introduction.\\

We want to sketch now how the missing arrow from Fig.~\ref{fig:mainDiagram} can be constructed, after resolving of inconsistencies of Section \ref{ss:problems}. First, it has to be understood how the transformations of the weighted bipartite graph on torus, corresponding to the mutations in $\mathcal{X}$-cluster algebra, should be properly uplifted to the transformation of quasi-periodically bipartite graph on a plane. Then, in the theory of total positivity, many of $A$-cluster variables are come as minors of the transfer matrices of paths on the bipartite graphs \cite{FZ98}, \cite{BFZ03}, \cite{P06}, or equivalently to the different minors of the Kasteleyn operator of this graph. We can relate then the different minors of infinite-dimensional $q$-difference Kasteleyn operator to the different $A$-cluster variables in deautonomized case. These minors also correspond to the partition functions of dimers with the different boundary conditions. Those, which are related by the unbounded ``rotations'' from Fig.~\ref{fig:gridTodaWithDimers2}, left, in the boxcounting limit present the same partition functions, but with the slightly shifted parameters. In our example, one can produce four different partition functions in this way, corresponding to $(Q_B,Q_F)$ and its shifts
\begin{equation}
(Q_{1,B},Q_{1,F}) \mapsto (q Q_{1,B},Q_{1,F}), ~~~
(Q_{1,B},Q_{1,F}) \mapsto (Q_{1,B}, q Q_{1,F}), ~~~
(Q_{1,B},Q_{1,F}) \mapsto (q Q_{1,B},q Q_{1,F}),
\end{equation}
which reproduces shifts of parameters in four $\tau$-functions in \cite{BGM17}. Then, the $q$-difference equations, satisfied by the dual topological string amplitudes become a Plucker relations between the regularized infinite dimensional minors of Kasteleyn operator, or exchange relations in the corresponding $A$-cluster algebra. The evidences of proper combinatorics, underlying this problem, might be contained in \cite{GKZ90}, \cite{S07}, \cite{F15}.\\

There is also a number of other intriguing directions, in which the developments of this paper might be continued:

\begin{itemize}
\item It is conjectured that all the fluctuations of the height function above the limit shape at ``infinite volume'' $q\to 1$ limit can be described using the Gaussian free field in the properly chosen complex structure, see e.g. \cite{K04}. In Section \ref{ss:height} using the quasi-classical computation for the zero-mode of Kasteleyn operator we provided a heuristic derivation for the height function of the limit shape. Similar quasi-classical computation for the Green function (\ref{eq:GreenSchematic}) would provide a solution for a problem of uniformization of fluctuations in spirit of \cite{KO05}: for any bipartite lattice and boundary conditions.
\item The distinguishing property of prepotential $\prep (U,Z)$ is that it satisfies the Seiberg-Witten equation (\ref{eq:SWeq}). However, this equation does not fix $Z$-dependence completely. There are also the so-called residue formulas and WDVV equations, which are differential equations on prepotential, involving $\partial/\partial Z$ derivatives \cite{M95}, \cite{GM13}. These formulas would be important approbations for prepotential (\ref{eq:prepdef}) as for the physical prepotential related to gauge theory.

The formula (\ref{eq:prepdef}) has to be extended also beyond the Harnak locus, since it essentially uses the property that the complex curve $P(e^z,e^w)=0$ projects 2 to 1 inside its amoeba. Another promising direction of studies is their extension to the case $P\neq Q$. This is a completely novel direction with no known analogue of Seiberg-Witten equation.
\item In \cite{BGM17} the quantization of cluster algebras \cite{BZ04}, \cite{FG03-2} was also applied, and the non-commutative $q$-difference bilinear equation on quantum $\tau$-functions where derived there as a result of application of several mutations. The solutions of these equations were provided there in terms of $5d$ Nekrasov functions with the generic $\Omega$-background, which generalizes the self-dual background of the commutative case. Our approach can be also generalized to this case in a straightforward way, promoting the face variables to be $t$-commutative, and performing the proper normal ordering. In this case, we expect the boxcounting formulas to be upgraded to the $(q,t)$ counting of ``refined topological vertices'' \cite{IKV07}. Similar ideas were proposed in \cite{N05}. Also the property of refined topological amplitude to intertwine the action of quantum toroidal algebra \cite{AFW11} might find its ``cluster'' interpretation using two-parametric quantization of classical $r$-matrix of \cite{GSV06}. It would be also interesting to ``refine'' results of \cite{DP22} in this setting.
\item The dimer models are similar to the Hermitian matrix models, since both can be described as specifications of Schur processes \cite{OR01}, \cite{MM22}. One of the most fundamental properties of matrix models is the genus expansion, when the diagrams of perturbation theory are interpreted as ribbon graphs, and the entire series is interpreted as a summation over all topologies. Similar expansion in $q$-case is more tricky and there is no final answer what to count as ``expansion over genuses'' in that case yet \cite{MPS20}. However, the dimer models might shed some light on this.

By bipartite graph on surface one can construct bipartite graph on dual surface by twisting all of its ribbons \cite{GK11}. This can be also done with the graph $\Gamma$ on the plane $\mathbb{R}^2$, getting the graph $\tilde{\Gamma}$ on the infinite genus, but ``regular'', dual surface $\tilde{S}$. Uplifting the paths, which are contributions to the normalized partition function of dimers, to the dual surface, one gets the set of cycles of non-trivial topology on $\tilde{S}$. Shrinking all the cycles on $\tilde{S}$, which are not winded by these paths, one gets finite genus curve, so the entire partition function becomes a summation over the surfaces of different topologies.

Once the expansion is properly formulated, one can find the observables for $q$-deformed resolvent, cut and joint, and check operators to obtain the loop equations and formulate $q$-topological recursion. This topological recursion might be also useful for the enumerative problems of \cite{JWY20} and  \cite{DP22}.

\item The phase space of cluster integrable system, as $\mathcal{X}$-cluster variety, is equipped with the logarithmically quadratic Poisson bracket for the face variables. For our main example from Fig.~\ref{fig:todanetwork} the quiver encoding this bracket is drawn on Fig.~2 from \cite{BGM17} under the name $A_7^{(1)'}$. The same quiver can be obtained\footnote{We are grateful to Fabrizio Del Monte for bringing our attention to this correspondence} by computing the Euler form of sheaves from the exceptional collection
\begin{equation}
\mathcal{C} = (\mathcal{O}(0),\mathcal{O}(1,0),\mathcal{O}(1,1),\mathcal{O}(2,1))
\end{equation}
of coherent sheaves on Hirzebruch surface $\mathbb{F}_0 = \mathbb{P}^1 \times \mathbb{P}^1$ \cite{BMP20}. More striking coincidence is that the formula (4.22) from \cite{BMP20} for the Chern classes $[N ;(c_{1,1},c_{1,2}); c_2]$ of the dual objects
\begin{equation}
\label{eq:todaClusterfByExcept}
\gamma_1 = [1; (0,0); 0],~~~ \gamma_2 = [-1; (1,0); 0], ~~~ \gamma_3 = [-1; (-1,1); 1],~~~ \gamma_4 = [1; (0,-1); 0]
\end{equation}
can be reproduced taking the ``finite'', not depending on $N$ and $M$ parts of degrees of $Q_i$ variables in (\ref{eq:todaClustersByQ}), and under identifications
\begin{equation}
\gamma_1 \leftrightarrow x_2, ~~~
\gamma_2 \leftrightarrow x_3, ~~~
\gamma_3 \leftrightarrow x_4, ~~~
\gamma_4 \leftrightarrow x_1,
\end{equation}
\begin{equation}
N \leftrightarrow \mathrm{deg}\, Q_2, ~~~
c_{1,1} \leftrightarrow \mathrm{deg}\, Q_{1,B}, ~~~
c_{1,2} \leftrightarrow \mathrm{deg}\, Q_{1,F}, ~~~
c_2 \leftrightarrow \mathrm{deg}\, Q_{0}.
\end{equation}
The correspondences above are precise to be just coincidence, so the dimer statistical model should have the deeper meaning in the counting of geometric objects, and there is a point to start. The local $3d$ Calabi-Yau, a mirror dual to the one defined by $uw = P(\lambda,\mu)$ with $P$ from (\ref{eq:LaurentPoly}), is the total space of the canonical bundle over $\mathbb{F}_0$ \cite{AHK97}, and $D$-branes on this total space are in correspondence with the exceptional collection of sheaves on the base \cite{BMP20}. And there is a straightforward way to produce more examples of this kind for check, since the both sides (either local $3d$ CY and cluster integrable system with the spectral curve  $P$) can be conveniently constructed starting from the Newton polygon.
\end{itemize}

\section*{Acknowledgments}
I am highly grateful to M.~Bershtein, P.~Gavrylenko and A.~Marshakov for our lengthy and fruitful conversations, which made this project possible. I want to thank also  F.~Del Monte, O.~Gamayun, A.~Grekov, R.~Gonin, N.~Iorgov, I. Krichever, A.~Okounkov, A.~Shapiro, A.~Shchechkin, I.~Vilkovisky and Y.~Zenkevich for numerous stimulating discussions. I am grateful to S.~Semenyakin for the careful proofreading of this manuscript. I want to thank to defenders of Ukraine, whose bravery gives me a hope for a prosperous future for my motherland.

\end{document}